%% file: main.tex
\documentclass[lettersize,journal]{IEEEtran}
\usepackage{amsmath,amsfonts}
\usepackage{algorithmic}
\usepackage{array}
\usepackage[caption=false,font=normalsize,labelfont=sf,textfont=sf]{subfig}
\usepackage{textcomp}
\usepackage{stfloats}
\usepackage{url}
\usepackage{verbatim}
\usepackage{graphicx}
\def\BibTeX{{\rm B\kern-.05em{\sc i\kern-.025em b}\kern-.08em
    T\kern-.1667em\lower.7ex\hbox{E}\kern-.125emX}}
\usepackage{balance}

\input{includes/includes}
\input{includes/macros}

\input{includes/pygments}
\usepackage[turnon]{notes}

\usepackage{soul} %
\sethlcolor{gray} 

\def\bf{\textbf}

\def\ie{\textit{i.e.,}}
\def\eg{\textit{e.g.,}}

\usepackage{balance}
\newcounter{o}
\usepackage{tikz}
\usepackage{eso-pic}
\usepackage{url}

\newcommand\copyrighttext{%
  \footnotesize This work has been submitted to the IEEE for possible publication. 
  Copyright may be transferred without notice, after which this version may no longer be accessible.\\
  Link: \url{https://uni.hi.is/helmut/2025/01/29/adding-copyright-information-to-submission-for-arxiv/}
}

\newcommand\copyrightnotice{%
  \AddToShipoutPictureBG*{%
    \AtPageLowerLeft{%
      \raisebox{2.2em}{\parbox{\textwidth}{\centering \copyrighttext}}
    }
  }
}

\begin{document}
\copyrightnotice
\title{LLM For Loop Invariant Generation and Fixing: How Far Are We?}

\author{
Mostafijur Rahman Akhond, 
Saikat Chakraborty  
and Gias Uddin
\thanks{Mostafijur Rahman Akhond and Gias Uddin are with York University, Canada (e-mail: mostafij@yorku.ca; guddin@yorku.ca).}
\thanks{Saikat Chakraborty is with Microsoft Research, USA (e-mail: saikatc@microsoft.com).}
}



\maketitle

\input{body/0.abstract}

\begin{IEEEkeywords}
Loop Invariants, LLM, GPT-4o, Formal Verification, SMT.
\end{IEEEkeywords}


\input{body/1.introduction}

\input{body/2.background}

\input{body/Methods}
\input{body/RQ1}

\input{body/RQ2}

\input{body/5.discussion}


\input{body/7.related}

\input{body/8.conclusion}



\bibliographystyle{IEEEtran}
\bibliography{main}



\end{document}

%% file: includes/includes.tex
\usepackage{arydshln}
\usepackage[font=small,labelfont=bf,tableposition=top]{caption}
\usepackage{hyperref}
\usepackage{boldline}
\usepackage{color,colortbl}
\usepackage{bigstrut}
\usepackage[ruled]{algorithm2e}
\usepackage{amsmath,amsfonts}
\usepackage{amsmath}
\usepackage{array}
\usepackage{algorithmic}
\usepackage{balance}
\usepackage{blindtext}
\usepackage{booktabs}
\usepackage{caption}
\usepackage{hyperref}
\usepackage[noabbrev]{cleveref}
\usepackage{comment}
\usepackage{enumitem}
\usepackage{eqparbox}
\usepackage{fancybox}
\usepackage{fancyvrb}
\usepackage{framed}
\usepackage{graphicx}
\usepackage{ifthen}
\usepackage{listings}
\usepackage{mathrsfs}
\usepackage{mdwmath}
\usepackage{mdwtab}
\usepackage{multirow}
\usepackage{pifont}
\usepackage{stfloats}
\usepackage{textcomp}
\usepackage{url}
\usepackage{xcolor}
\usepackage{xspace}
\usepackage[normalem]{ulem}
\usepackage[framemethod=TikZ]{mdframed}
\usepackage{caption}
\usepackage{subcaption}
\usepackage{tabularx}
\usepackage{titlesec}
\usepackage{tcolorbox}
\tcbuselibrary{listings,skins}
\usepackage{mathtools}
\usepackage{blindtext}
\usepackage{lstautogobble}

%% file: includes/macros.tex
\usepackage{pifont}

\usepackage{xspace}
\xspace%

\usepackage{tcolorbox}

\newtcbox{\inlinebox}[1][]{enhanced,
 box align=base,
 nobeforeafter,
 colback=blueish,
 size=small,
 left=0pt,
 right=0pt,
 boxsep=2pt,
 #1}

\newcommand{\lessons}[1]{
    \begin{lesson}
    \small
       O#1
    \end{lesson}
}

\usepackage{cleveref}

\renewcommand{\cref}[1]{\Cref{#1}}

\newcommand{\rom}[1]{\uppercase\expandafter{\romannumeral #1\relax}}

\newcommand{\etal}{\hbox{\emph{et al.}}\xspace}
\newcommand{\eg}{\hbox{\emph{e.g.,}}\xspace}
\newcommand{\ie}{\hbox{\emph{i.e.,}}\xspace}

\definecolor{gray50}{gray}{.5}
\definecolor{gray40}{gray}{.6}
\definecolor{gray30}{gray}{.7}
\definecolor{gray20}{gray}{.8}
\definecolor{gray10}{gray}{.9}
\definecolor{gray05}{gray}{.95}
\definecolor{gray01}{gray}{.97}

\newlength\Linewidth
\def\findlength{\setlength\Linewidth\linewidth
\addtolength\Linewidth{-4\fboxrule}
\addtolength\Linewidth{-3\fboxsep}
}
\newenvironment{examplebox}{\par\begingroup
  \setlength{\fboxsep}{3pt}\findlength
  \setbox0=\vbox\bgroup\noindent
  \hsize=0.90\linewidth
  \begin{minipage}{0.90\linewidth}\normalsize}
    {\end{minipage}\egroup
    \textcolor{gray20}{\fboxsep1.2pt\fbox
     {\fboxsep5pt\colorbox{gray05}{\normalcolor\box0}}}
    \endgroup\par\noindent
    \normalcolor\ignorespacesafterend}

\usepackage{tikz}
\usetikzlibrary{arrows.meta, shapes, positioning, calc}
\usepackage{mdframed}

\usetikzlibrary{shadows}
\usepackage{graphics}
\newmdenv[
    tikzsetting= {fill=blueish},
    skipabove=0.33em,
    skipbelow=0.33em,
    linewidth=1pt,
    innerleftmargin=4pt,
    innerrightmargin=4pt,
    innertopmargin=2pt,
    innerbottommargin=2pt,
    linecolor=gray95,
    roundcorner=2pt, 
    shadow=true,
    shadowsize=4pt,
    shadowcolor=gray95
]{questionbox}

\newmdenv[
    tikzsetting= {fill=greenish},
    skipabove=0.33em,
    skipbelow=0.33em,
    linewidth=1pt,
    innerleftmargin=4pt,
    innerrightmargin=4pt,
    innertopmargin=2pt,
    innerbottommargin=2pt,
    linecolor=gray95,
    roundcorner=2pt, 
    shadow=true,
    shadowsize=4pt,
    shadowcolor=gray95
]{answerbox}

\newmdenv[
    skipabove=0.33em,
    skipbelow=0.33em,
    innerleftmargin=4pt,
    innerrightmargin=4pt,
    innertopmargin=2pt,
    innerbottommargin=2pt,
]{lessonbox}

\usepackage{tikz}

\usepackage{tabularx}

\newenvironment{lesson}
{
    \begin{lessonbox}
}
{\end{lessonbox}}

\newenvironment{result}
{\begin{answerbox}}
{\end{answerbox}}

\newenvironment{question}
{\begin{questionbox}}
{\end{questionbox}}

\definecolor{javared}{rgb}{0.6,0,0} 
\definecolor{javagreen}{rgb}{0.25,0.5,0.35} 
\definecolor{javapurple}{rgb}{0.5,0,0.35} 
\definecolor{javadocblue}{rgb}{0.25,0.35,0.75} 

\lstdefinestyle{basejava}{
  language=java,
  showstringspaces=false,
  basicstyle=\scriptsize\ttfamily,
  keywordstyle=\bfseries\color{javapurple},
  commentstyle=\itshape\blue,
  identifierstyle=\blue,
  frame=none,
  backgroundcolor=\color{white},
}

\lstdefinestyle{CustomJava}{
  belowcaptionskip=\baselineskip,
  breaklines=true,
  xleftmargin=\parindent,
  language=java,
  showstringspaces=false,
  basicstyle=\scriptsize\ttfamily,
  keywordstyle=\bfseries\color{javapurple},
  commentstyle=\itshape\blue,
  identifierstyle=\blue,
  belowskip=1pt,
  frame=shadowbox,
  backgroundcolor=\color{gray01},
  gobble=0
}

\usepackage{tcolorbox}
\tcbuselibrary{listingsutf8}  

\newtcblisting{mylisting}[2][]{
    arc=3mm,
    listing only, 
    listing style=codit,
    title=#2,
    #1
    }

%% file: includes/pygments.tex
\newcommand\blue[1]{\textcolor[rgb]{0.00,0.00,1.00}{{#1}}}

\definecolor{blueish}{RGB}{252, 252, 255}
\definecolor{greenish}{RGB}{252, 255, 252}
\definecolor{redish}{RGB}{255, 250, 250}

\definecolor{gray05}{gray}{0.95}
\definecolor{gray08}{gray}{0.92}
\definecolor{gray10}{gray}{0.90}
\definecolor{gray12}{gray}{0.88}
\definecolor{gray15}{gray}{0.85}
\definecolor{gray18}{gray}{0.82}
\definecolor{gray20}{gray}{0.80}
\definecolor{gray25}{gray}{0.75}
\definecolor{gray30}{gray}{0.70}
\definecolor{gray35}{gray}{0.65}
\definecolor{gray40}{gray}{0.60}
\definecolor{gray45}{gray}{0.55}
\definecolor{gray50}{gray}{0.50}
\definecolor{gray55}{gray}{0.45}
\definecolor{gray60}{gray}{0.40}
\definecolor{gray65}{gray}{0.35}
\definecolor{gray70}{gray}{0.30}
\definecolor{gray75}{gray}{0.25}
\definecolor{gray80}{gray}{0.20}
\definecolor{gray85}{gray}{0.15}
\definecolor{gray90}{gray}{0.10}
\definecolor{gray95}{gray}{0.05}

\definecolor{fstarid}{rgb}{0.28,0.07,0.07}

\definecolor{addition}{rgb}{0,0.1,0.5}
\definecolor{dkblue}{rgb}{0,0.0,0.7}
\definecolor{dkgreen}{rgb}{0,0.4,0}
\definecolor{dkred}{rgb}{0.6,0,0}
\definecolor{dkpurple}{rgb}{0.55,0,0.75}
\definecolor{purple}{rgb}{0.69,0,.87}
\definecolor{olive}{rgb}{0.4, 0.4, 0.0}
\definecolor{teal}{rgb}{0.0,0.4,0.4}
\definecolor{azure}{rgb}{0.0, 0.5, 1.0}
\definecolor{gray}{rgb}{0.5, 0.5, 0.5}
\definecolor{dkgrey}{rgb}{0.2, 0.2, 0.2}
\definecolor{lilac}{rgb}{0.70, 0.04, 0.08}
\definecolor{applegreen}{rgb}{0.55, 0.71, 0.0}

\definecolor{step1}{HTML}{CC0066}
\definecolor{step2}{HTML}{CC6600}
\definecolor{step3}{HTML}{440077}
\definecolor{step4}{HTML}{007700}
\definecolor{step5}{HTML}{0000FF}

%% file: body/0.abstract.tex
\begin{abstract}   
    A loop invariant is a property of a loop that remains true before and after each execution of the loop. The identification of loop invariants is a critical step to support automated program safety assessment. 
    Recent advancements in Large Language Models (LLMs) have demonstrated potential in diverse software engineering (SE) and formal verification tasks. However, we are not aware of the performance of LLMs to infer loop invariants.
    We report an empirical study of both open-source and closed-source LLMs of varying sizes to assess their proficiency in inferring inductive loop invariants for programs and in fixing incorrect invariants. 
    Our findings reveal that while LLMs exhibit some utility in inferring and repairing loop invariants, their performance is substantially enhanced when supplemented with auxiliary information such as domain knowledge and illustrative examples. LLMs achieve a maximum success rate of 78\% in generating, but are limited to 16\% in repairing the invariant.    
    
    

\end{abstract}

%% file: body/1.introduction.tex

\section{Introduction}
\IEEEPARstart{S}{ynthesizing} loop invariants is a fundamental problem in software verification~\cite{karr1976affine, furia2014loop, si2018learning}. A loop invariant is a logical assertion that remains true before and after each iteration of a loop and is essential for proving program correctness without execution~\cite{colon2003linear, chakraborty2023ranking}. Due to the undecidability of general program behavior involving loops~\cite{hrushovski2023strongest, muller2004computing}, generating loop invariants is inherently undecidable in the general case. Over the years, researchers have proposed various approaches, each with distinct strengths and limitations. Among the earliest methods, static analysis~\cite{karr1976affine} and symbolic execution techniques~\cite{flanagan2001houdini, corr18padhi, lahiri2007predicate} have shown considerable success, especially in the past decade. However, these techniques often struggle to scale or adapt to modern and more complex programming constructs, limiting their effectiveness in newer verification scenarios.



With the advent of machine learning (ML), several models have been developed for loop invariant synthesis. Among them, Padhi \etal~\cite{padhi2016data} introduced a data-driven approach that learns from counterexamples and generalizes to unseen problems. Another notable contribution by Si \etal~\cite{si2018learning} applied reinforcement learning to the task. However, many of these ML-based methods face challenges when applied to programs with complex loop structures. More recently, large language models (LLMs) have shown promise in this domain, leveraging their general-purpose generative capabilities. Chakraborty \etal~\cite{chakraborty2023ranking} demonstrated that LLMs can generate loop invariants for Satisfiability-Modulo-Theories (SMT) program specifications and proposed a ranking model to identify correct ones efficiently. Kamath \etal~\cite{kamath2023finding} introduced an oracle-guided prompting technique for Frama-C loops, while Misu \etal~\cite{misu2024towards} proposed a few-shot prompting strategy for verified method generation in Dafny. These efforts highlight the potential of LLMs for invariant generation, though a comprehensive evaluation of their optimal configurations remains an open research question.



In this research, we empirically investigate how large LLMs reason through the loop invariant synthesis process, especially for SMT specifications. We answer two major research questions (RQs):
\begin{enumerate}[label=\textbf{RQ\arabic{*}.}, leftmargin=20pt]
    \item \textbf{Generation ability of LLM:}  
    How good are LLMs at generating loop invariants from
program specifications?
     \item \textbf{Repair ability of LLM:}  Can LLM repair the incorrect loop invariants?
\end{enumerate}

Prior work has demonstrated that LLM performance improves significantly when guided by well-structured instructions \cite{zeng2023evaluating}. One effective strategy involves supplying relevant background knowledge. In our study, we provided such knowledge in two forms: (1) a set of instructions outlining the nature of the problem and the steps for synthesizing an invariant, and (2) problem decomposition, with breaking a complex task into smaller, more manageable subproblems and synthesizing partial solutions in a divide-and-conquer fashion. Problem partitioning has been shown to aid LLM performance in other domains~\cite{wu2024divide}, though its effectiveness in invariant synthesis remains unexplored. Another widely adopted LLM instruction setup is few-shot prompting \cite{reynolds2021prompt}; however, its efficacy in loop invariant synthesis has yet to be conclusively demonstrated. We experimented with multiple example configurations to identify the most effective setup. We also tested a combined approach that fuses background knowledge with few-shot prompts. The results in this study reveal several important insights about the LLMs' capabilities in synthesizing loop invariants. LLM achieves up to 49\% success in generating verified invariants by following a series of guiding instructions. However, with a few-shot prompting approach, the success rate increases to 76\%. Though a mix of guiding instructions and a few-shot prompting performs marginally improved to 78\%, its applicability is limited due the the higher size of input tokens.

Both the literature \cite{chakraborty2023ranking} and our findings suggest that LLMs often require a large number of samples to produce a correct invariant, raising concerns about their efficiency and requiring further investigation into their ability to repair incorrect invariants. Fortunately, verifiers like Z3~\cite{de2008z3} provide rich feedback, including the logical condition that failed and the specific variable assignments causing the failure. Motivated by this, we evaluated whether such verifier-provided feedback could assist LLMs in repairing incorrect invariants. We examined two repair scenarios: one where the LLM was given the general cause of failure, and another where it was provided with the concrete variable values responsible for the failure. However, LLMs showed limited ability to leverage this information effectively, achieving up to 16\% accuracy in repairing failed invariants.

%% file: body/2.background.tex
\section{Background} \label{ssec: background}




Let us assume a simple grammar, $\mathcal{G}$, of a programming language that enables writing program statements $S$.
The central objective of program verification is to ensure that a given program behaves according to its specification, typically expressed as a \emph{Hoare triple}~\cite{hoare1969axiomatic}:
\begin{equation}
    \{ \phi (v) \}~p~\{ \psi (v) \}
    \label{eqn:generic_inv}
\end{equation}

Where $p$ is the program in ${\mathcal{G}}$, ${v}$ is a set of variables representing the program state. \Cref{eqn:generic_inv} asserts that for any terminating execution of $p$, if the precondition $\phi(v)$ holds in the initial state, then the postcondition $\psi(v)$ must hold in the final state.
When $\mathcal{G}$ is extended to allow program that includes loops controlled by a Boolean expression $b$, verifying correctness requires identifying a suitable {\emph{loop invariant} $I(v)$}. For a loop of the form {\texttt{while}~$b$~\texttt{do}~$S$}, a valid invariant $I(v)$ must satisfy the following conditions:

{
    \begin{equation}
        \begin{aligned}
            \{ \phi(v) \}~\texttt{skip}~\{ I(v) \} \\
            \{ I(v_{old}) \wedge b \}~S~\{ I(v_{new}) \}  \\
            \{ I(v) \wedge \neg b \}~\texttt{skip}~\{ \psi(v) \}
        \end{aligned}
        \label{eq:loop-conditions}
    \end{equation}
}
These conditions denote that the invariant $I(v)$ holds true at the loop entry, is preserved at each iteration of the loop, and satisfies the postcondition upon the termination of the loop. The problem of \emph{loop invariant inference} is to synthesize an $I$ that satisfies all three conditions in~\eqref{eq:loop-conditions}, which we denote as $I \vdash p$.

\begin{figure*}[t]
    \centering
    \includegraphics[width=0.9\linewidth]{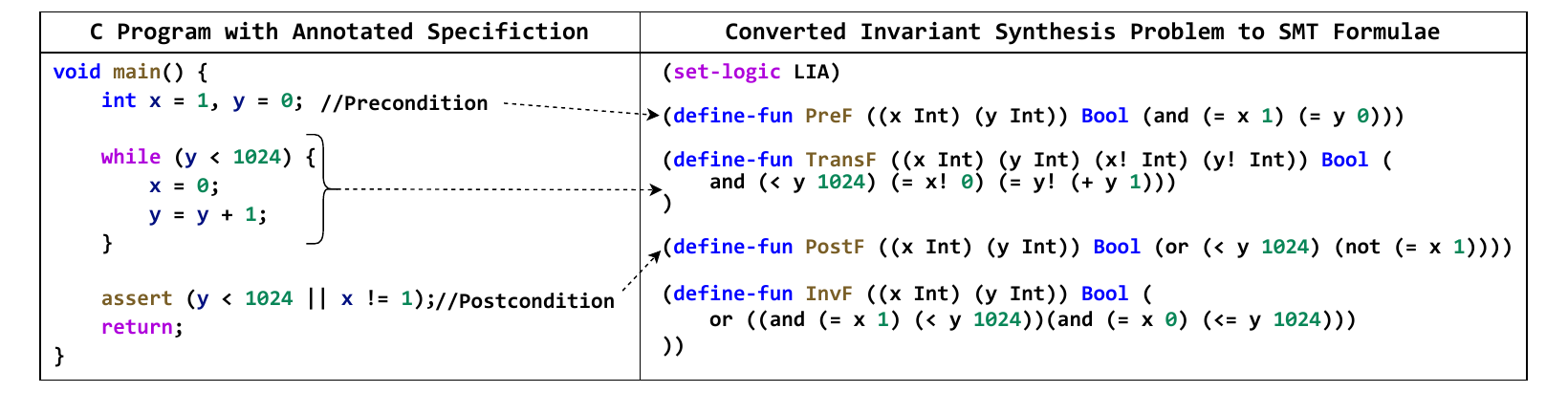}
    \caption{An Example C program and its translation to SMT formulae. Here, $x!$ and $y!$ denote the updated state of variable $x$ and $y$, respectively, after each iteration of the loop. The \texttt{PreF} denotes the preconditions, \texttt{PostF} denotes the postcondition, and the \texttt{TransF} denotes the summary of the loop. The \texttt{InvF} is the inductive loop invariant that can verify that the program follows these specifications. }
    \label{fig:loopInvExample}
\end{figure*}



\begin{align}
   PreF(x, y) \Rightarrow InvF(x, y) \tag{R1} \label{c2} \\
   InvF(x, y) \land TransF(x, y, x\text{\small !}, y\text{\small !}) \Rightarrow InvF(x\text{\small !}, y\text{\small !}) \tag{R2} \label{c3} \\
  InvF(x, y) \Rightarrow PostF(x, y)\tag{R3} \label{c4}
\end{align}

Given a program loop as input, the task of loop invariant synthesis is to deterministically translate the program into a set of logical functions that offer specifications for the precondition and postcondition for the loop. 
Figure~\ref{fig:loopInvExample} illustrates an example C program for which we show the specifications in SMT representations. 
Three key functions ($PreF$, $TransF$, and $PostF$) are used to define the invariant $InvF$. They are used to create three satisfiability rules - \ref{c2}, \ref{c3}, \& \ref{c4} (followed from \Cref{eq:loop-conditions}). The rule \ref{c2} corresponds to the first line of \Cref{eq:loop-conditions}, similarly \ref{c3}, \& \ref{c4} correspond to the second and third lines of \Cref{eq:loop-conditions}.  An automated theorem prover like Z3 can check if all three rules hold by the invariant. If any rule fails, the loop invariant is considered a failed invariant.

%% file: body/Methods.tex
\section{Study Methodology} 
Our goal is to study the effectiveness of LLMs to create and repair loop invariants. The subject of our study is a list of LLMs and the object is a dataset of loop invariants. Our two major research questions (RQs) are as follows.

\begin{enumerate}[label=\textbf{RQ\arabic{*}.}]
    \item How good are LLMs at generating loop invariants from program specifications?
    \item Can LLM repair the incorrect loop invariants?
\end{enumerate}
We analyze the loop invariant generation capability of LLMs (i.e., RQ1) by answering the following sub-RQs:
\begin{enumerate}[label=\textbf{RQ1.\arabic{*}.}, leftmargin=35pt]
     
    \item  Can an LLM synthesize loop invariants when given instructions with domain knowledge?
    
    \item  Can an LLM synthesize better loop invariants when the problem is partitioned by the three invariant conditions?
    
    \item  Can few-shot prompting with similar examples impact the invariant synthesis? 
    
    \item Can the type of the examples impact few-shot prompting for invariant synthesis?  
    
    \item  Can we combine domain knowledge with few-shot examples to synthesize more accurate loop invariant?
    
\end{enumerate}
We study the fixing capability of LLMs to repair faulty loop invariants (i.e., RQ2) by answering the following sub-RQs:
\begin{enumerate}[label=\textbf{RQ2.\arabic{*}.}, leftmargin=35pt]
    \item  Can feedback to an LLM help it repair
an invariant that it generated incorrectly?
    \item Does providing information about failure root causes and logs help an LLM repair an incorrect invariant?
\end{enumerate}




    




\begin{figure}[!t]
    \centering
    \includegraphics[width=\linewidth]{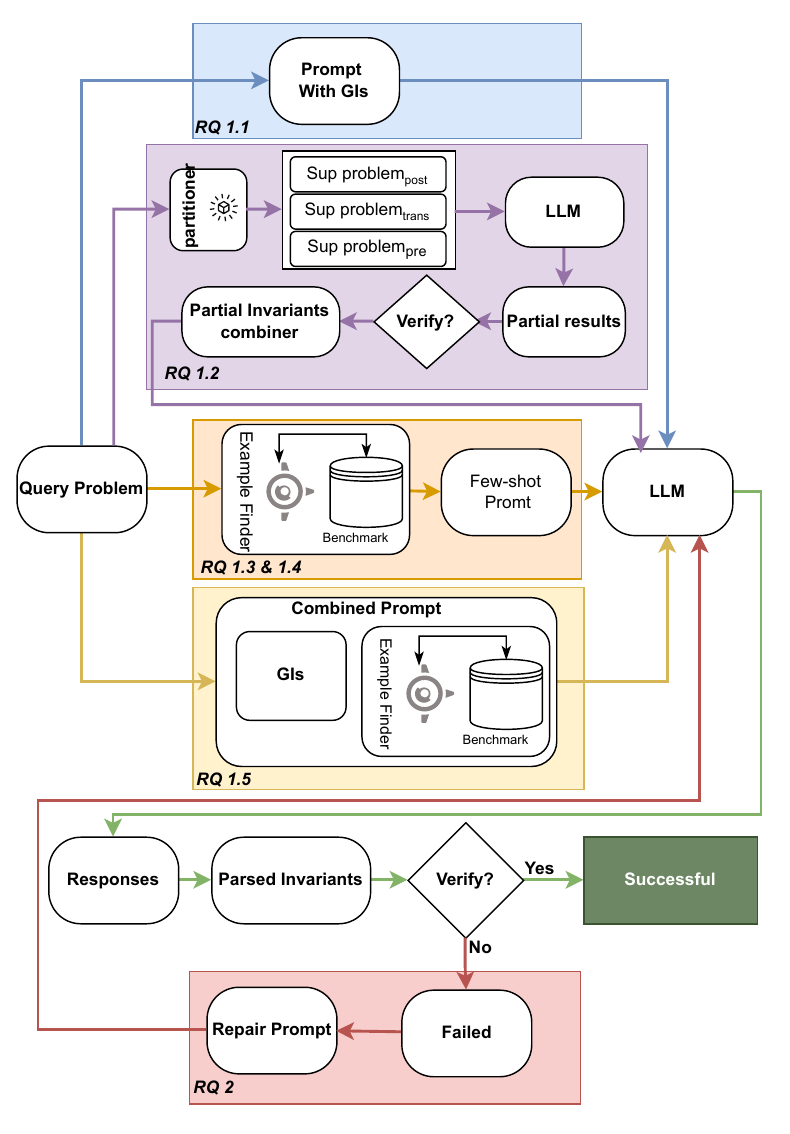}
    \caption{General workflow of the experiment}
    \label{fig:model}
\end{figure}

~\Cref{fig:model} illustrates the overall workflow of our experiments to answer the RQs.  The input is a query as an invariant synthesis problem $\mathcal{P}$ represented in SMT format. 
Based on the query, we construct a prompt using one of several prompt builders (shown as \textbf{RQ1.*} boxes in ~\cref{fig:model}) and send it to an LLM, which generates a list of candidate invariants. 
An invariant parser extracts these invariants from the LLM outputs. Each invariant is checked by a Z3 verifier for correctness. 
If the invariant is incorrect, we create a repair prompt using the verifier’s feedback and ask the LLM to revise the invariant. 
We re-parse and re-verify the revised output. 
This cycle continues until we find a valid invariant or we reach a selected number of iterations.

\noindent\textbf{Studied Dataset.} We utilized the loop invariant synthesis benchmark dataset curated by Padhi~\etal~ \cite{padhi2016data}. This dataset consists of 940 problems formatted in SMT~\cite{alur2013syntax}. The dataset was further analyzed by Chakraborty~\etal~\cite{chakraborty2023ranking}, who identified 429 problems with at least one invariant satisfying all verification conditions. From these 429 problems, we selected a representative subset of 210 problems 
using random sampling, guided by a 98\% confidence level with 5\% interval, and 75\% proportion size.  \cite{uddin2022empirical}.

\noindent\textbf{Studied LLMs.} We selected several well-established LLMs from both closed-source and open-source distributions. \Cref{tab:llm_models} provides a summary of the selected models along with their corresponding configurations. To set up the LLMs for the experiment, for Llama 3.1, we used Google Colab with an A100 GPU, equipped with 83GB of RAM and 40GB of GPU memory. For GPT-4 and Mistral-large, we accessed their official APIs.


\begin{table}[!t]
\caption{Overview of Studied LLM Models}
\label{tab:llm_models}
\footnotesize
\centering
\begin{tabular}{p{0.2\linewidth}|p{0.18\linewidth}|p{0.5\linewidth}}
\textbf{Model} & \textbf{\# Parameters} & \textbf{Overview} \\
\hlineB{2}
\textbf{Llama 3.1 (8B-Instruct)\cite{meta2024llama3} } & 8 Billion & Open source model, tested on locally deployed machine \\
\hline
\textbf{Mistral-large (Instruct-2407)\cite{mistral_large}} & 123 Billion & Open model for research, accessed through official API \\
\hline
\textbf{GPT-4o (2024-05-13)\cite{achiam2023gpt}} & 2 Trillion & High performance closed source model, accessed through official API \\
\hlineB{2}
\end{tabular}
\end{table}


%% file: body/RQ1.tex
\section{Loop Invariant Generation by LLMs (RQ1)}

\subsection{RQ1.1 Can an LLM synthesize loop invariants when given instructions with domain knowledge?}

\subsubsection{Motivation}

Large Language Models (LLMs) have demonstrated strong capabilities across a variety of tasks when given appropriate instructions~\cite{chang2024survey}. However, their effectiveness can be limited in domains that require deep reasoning or specialized knowledge, such as program analysis and invariant generation. It is necessary to understand whether LLMs can produce loop invariants when given domain-specific guidance~\cite{zeng2023evaluating} as simple instructions. 

\subsubsection{Approach}

\begin{figure}[t]
    \centering
    \includegraphics[width=\linewidth]{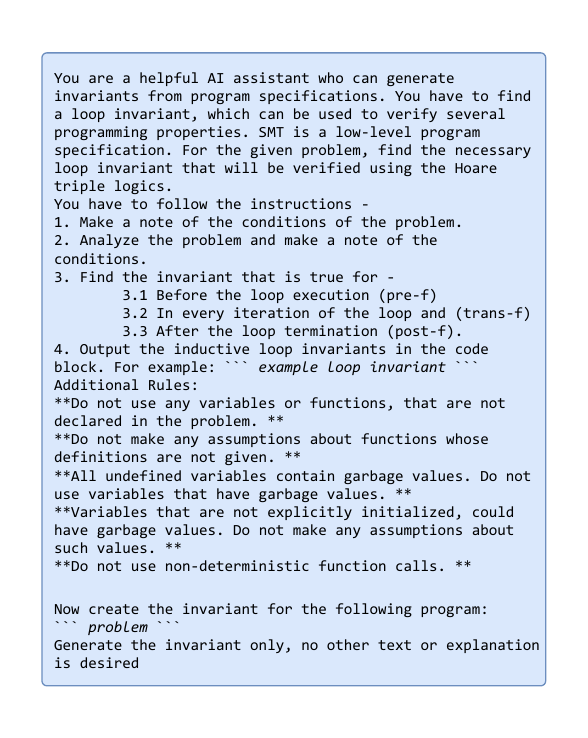}
    \caption{Prompt template designed to generate invariants using instructions with a large language model.}
    \label{fig:prompt_gi}
\end{figure}

We developed a set of instructions to guide the LLM models about the problem domain-specific knowledge. To formulate these instructions, we studied the invariant generation process from different state-of-the-art research \cite{padhi2016data, achiam2023gpt, chakraborty2023ranking, wu2023lemur, misu2024towards}, translated the process into structured guidelines, and organized them into a set of instructions for LLM. 
Figure \ref{fig:prompt_gi} illustrates a prompt template containing the Instructions. These instructions are composed of three main parts: 1) In the first part (first seven lines) of the template, we introduced the problem and defined LLM's role here. 2) The second part is presented as a numbered list of instructions. It guides about the key concerns to be considered while generating the invariants, and the logical consequences to maintain. 3) The third part defines additional rules that constrain LLMs to generate invariants using only the specified variables and functions.



\subsubsection{Result}

\begin{table}[!t]
\caption{Study on LLMs' Invariant Generation Capability}
\label{tab: GIcomparision}
\footnotesize
\centering

\begin{tabular}{l|l|l}
\textbf{LLM} & \textbf{\% Followed template}  & \textbf{\% Syntactically correct} \\
\hlineB{2}
\textbf{Llama 3.1 8b} & \multicolumn{1}{c|}{0} & \multicolumn{1}{c}{0} ~\bigstrut\\
\hline
\textbf{Mistral-large} & \multicolumn{1}{c|}{83.33} & \multicolumn{1}{c}{73.33} ~\bigstrut \\
\hline
\textbf{GPT-4o} & \multicolumn{1}{c|}{100} & \multicolumn{1}{c}{100} ~\bigstrut \\
\hlineB{2}
\end{tabular}
\end{table}

Table \ref{tab: GIcomparision} summarizes the experimental findings on the LLMs' ability to generate invariants from instructions.  
All three LLMs generated some responses, regardless of whether those responses contain to expected invariant or not.  
We observe that the responses of Llama 3.1 mostly contain natural language descriptions of the problem it is supposed to solve -- acting more like a helper to the developer who is writing the invariant. 
None of the generated responses by Llama-3.1 followed the output template we instructed in the prompt
As such, our automated invariant extraction tool is not able to extract {\em any} invariant for further verification and evaluation. 
There are a couple of instances in Llama-3.1 responses, where our best effort heuristic response parser extracted an invariant -- almost all of those contained syntactic error, mostly related to imbalanced parentheses. 
Mistral-large responses were much better, where 83.33\% of the solutions followed the output template, and 73.33\% of them were syntactically correct. 
All the responses from GPT-4o correctly followed the output template, and we did not encounter any syntax error while processing any of the automatically extracted invariants from these responses. 
\addtocounter{o}{1}\lessons{\theo. GPT-4o and Mistral-large generated syntactically correct loop invariants in 100\% and 73.33\%  of the cases, whereas Llama 3.1 (8B) failed to produce useful responses relevant to our study.}

\begin{table}[!t]
    \centering
    \caption{Performance comparison of LLM models on generating invariants following instructions.}
    \footnotesize
    
    { \begin{tabular}{c|c|c|c|c|c|c}
        & \multicolumn{6}{c}{\textbf{\% of problems solved in k generated responses}}\bigstrut \\
        \cline{2-7}
        \multirow{2}{*}{\textbf{LLM}} & \multicolumn{3}{c|}{\textbf{Without Instructions}} & \multicolumn{3}{c}{\textbf{With Instructions}}\bigstrut[t] \\
        & $k=10$ & $k=30$ & $k=50$ & $k=10$ & $k=30$ & $k=50$ \bigstrut[b] \\
        \hlineB{2}
        Mistral-large & 13 & 18 & 21 & 21 & 27 & 31  \bigstrut\\ 
        GPT-4o & 19 & 32 & 40 & 27 & 41 & 49 \bigstrut \\
        \hlineB{2}
        \end{tabular}
    }
    \label{tab:GI_performance}
\end{table}

Given the poor performance of Llama-3.1, we restrict our further investigation to only Mistral-large and GPT-4o.
To assess the effectiveness of the Instructions, we tasked GPT-4 and Mistral-large with generating 50 invariants ($k=50$) for each of the problems using the instruction-based prompt (\Cref{fig:prompt_gi}). 
As shown in ~\Cref{tab: GIcomparision}, both Mistral-large and GPT-4o are benefited from the instructions. 
GPT-4 generates at least one successful invariant for 49\% (103 out of 210) of the problems when we put the instructions in the prompt, compared to 40\% without the instructions.
We also observe a similar impact of instructions in Mistral-large. 
Note that we do not intend to compare performance across different LLMs, our goal is to empirically identify the impact of the instructions while generating loop invariants.
Experimentally, we observe that both models demonstrated a comparatively higher success rate when using the instruction-based prompts than when using prompts without those instructions. 
This highlights the positive impact of the instruction set on the models' performance in generating valid invariants.   
\addtocounter{o}{1}\lessons{\theo. Both Mistral-large and GPT-4o generated more successful invariants with the instructions, achieving success rates of 31\% and 49\% respectively for a sample size of $k = 50$.}

\subsection{RQ1.2 Can an LLM synthesize better loop invariants when the problem is partitioned by the three invariant  conditions?}

\subsubsection{Motivation} In RQ1.1, we provided a set of instructions to an LLM to produce an invariant that satisfies all three conditions of a loop invariant at once. Given the complexity of the loop invariants, it may be more effective to decompose the task, i.e., asking an LLM to produce rules per condition. 

\subsubsection{Approach} We approached the problem of loop invariant synthesis in a divide-and-conquer fashion. As such, we instructed an LLM separately for each invariant-related condition. 
After generating invariants for each of the conditions, we then asked the LLM to integrate so that it can satisfy all three conditions.

\begin{figure}[!t]
    \centering
    \includegraphics[width=\linewidth]{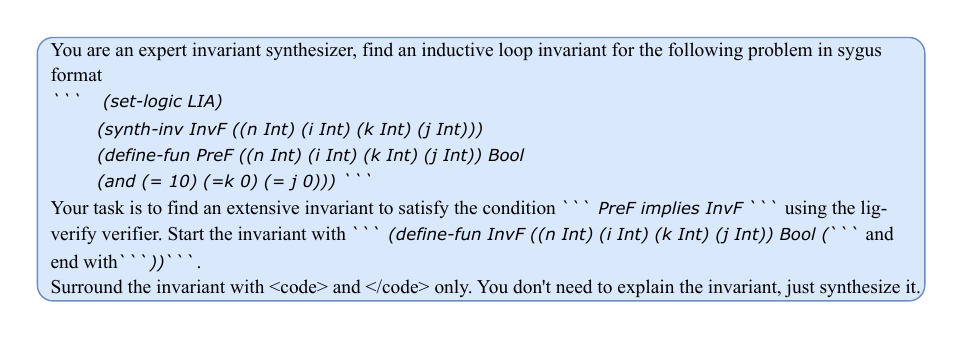}
    \caption{Prompt template to utilize partial invariant synthesis}
    \label{fig:partial-prompt}
\end{figure}

\begin{figure}[!t]
    \centering
    \includegraphics[width=\linewidth]{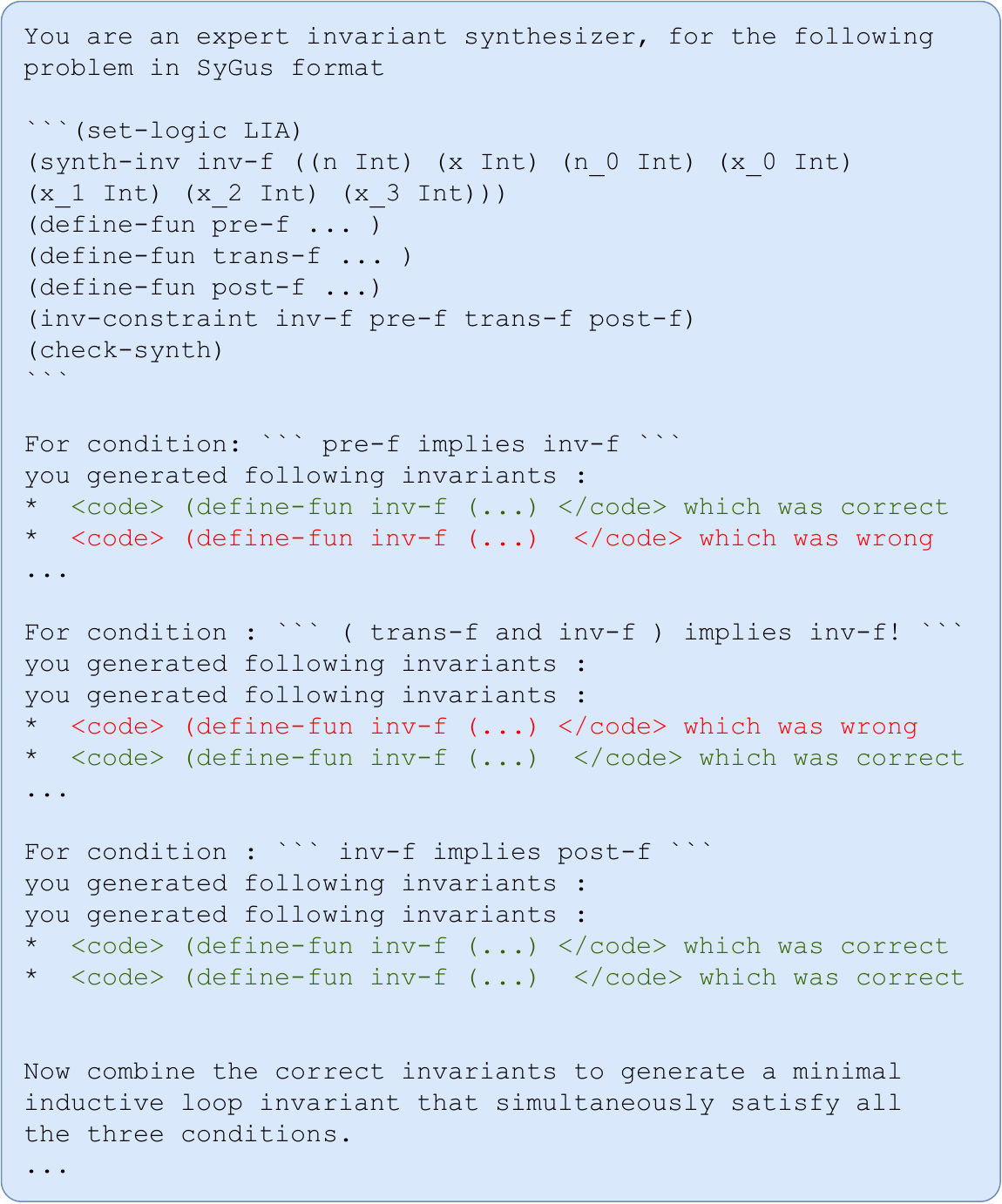}
    \caption{Prompt for combining partial invariants}
    \label{fig:combined_prompt}
\end{figure}

To set this experiment, we designed a new prompt (figure \ref{fig:partial-prompt}) for a single condition above, omitting the details of unrelated things. 
For example, loop satisfiability condition \ref{c2} requires precondition and invariant only; we omit details related to the loop body and postcondition from this prompt.   
In this way, we are relaxing the problem into three easier subproblems. We call these three prompts $\mathcal{P}_{pre}$, $\mathcal{P}_{trans}$, and $\mathcal{P}_{post}$ (referred collectively as $\mathcal{P}_{partial}$), respectively each corresponding to only the conditions \ref{c2}, \ref{c3}, and \ref{c4}, respectively. 
After getting the invariants ($\mathcal{I}_{conditional}$) for each condition, we asked LLMs to combine them to generate the final invariants($\mathcal{I}_{final}$) -- we call this prompt $\mathcal{P}_{full}$. 
Concretely, we instructed the LLMs to generate ten $\mathcal{I}_{conditional}$s ($k=10$) independently using each of $\mathcal{P}_{pre}$, $\mathcal{P}_{trans}$, and $\mathcal{P}_{post}$. 
We then test the generated $\mathcal{I}_{conditional}$ using a modified version of the verifier, which only checks satisfiability of the corresponding condition.
Thus, for each problem, we have thirty $\mathcal{I}_{conditional}$s and corresponding verification results.
Next, we create the $\mathcal{P}_{full}$ (see \Cref{fig:combined_prompt} as an illustration) with all of these $\mathcal{I}_{conditional}$s and corresponding results and instruct LLMs to generate fifty ($k=50$) $\mathcal{I}_{final}$. 

Our rationale behind showing the failed candidate from an individual trial in addition to the successful ones is that we want to demonstrate what had gone wrong (since we are also presenting the partial verification results), and help LLMs avoid certain mistakes. 

\begin{figure}[!t]
    \centering
    \includegraphics[width=\linewidth]{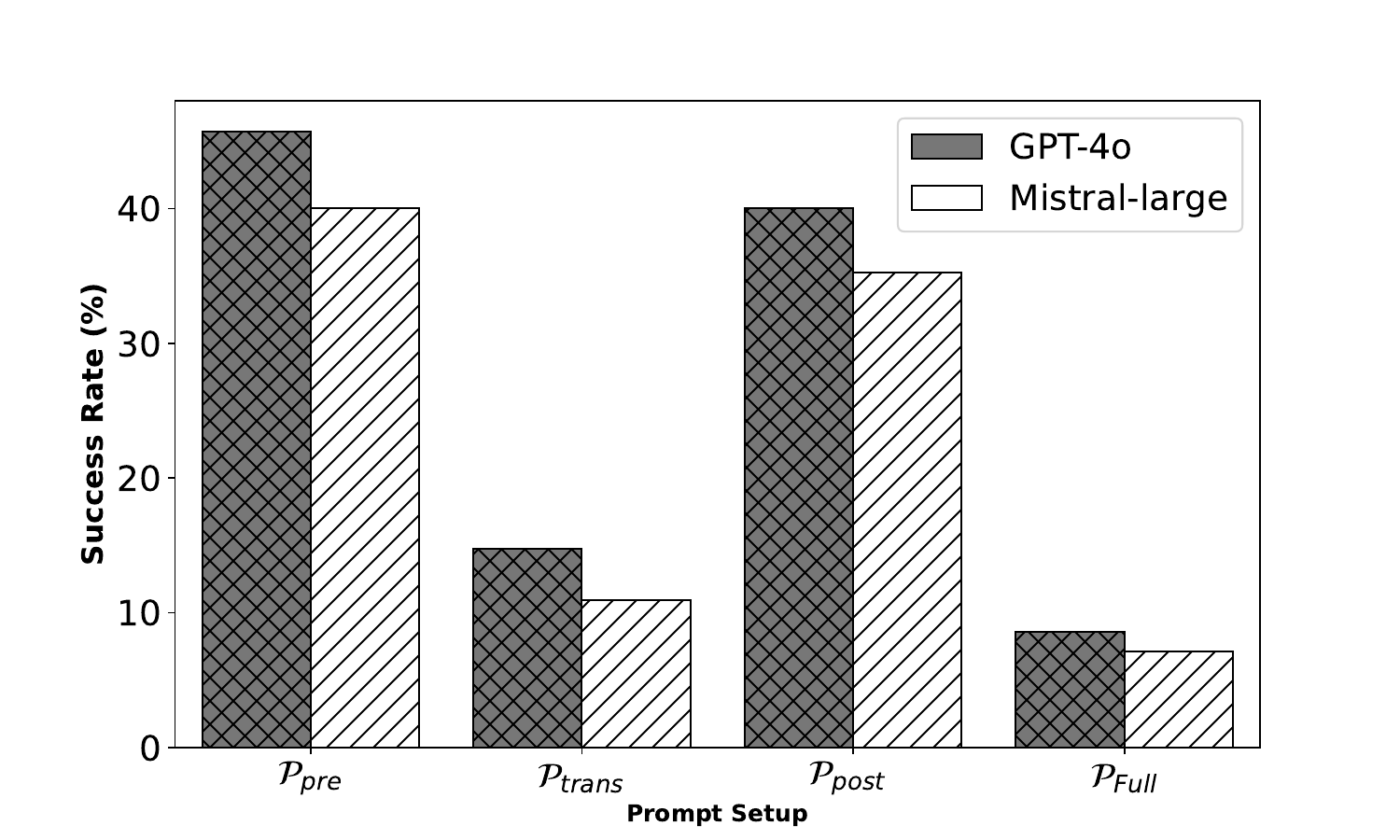}
    \caption{Success rate with partitioning invariant problem}
    \label{fig:partial-inv-res}
\end{figure}

\subsubsection{Result}


\begin{figure}[!t]
\centering
\subfloat[]{\includegraphics[width=2.5in]{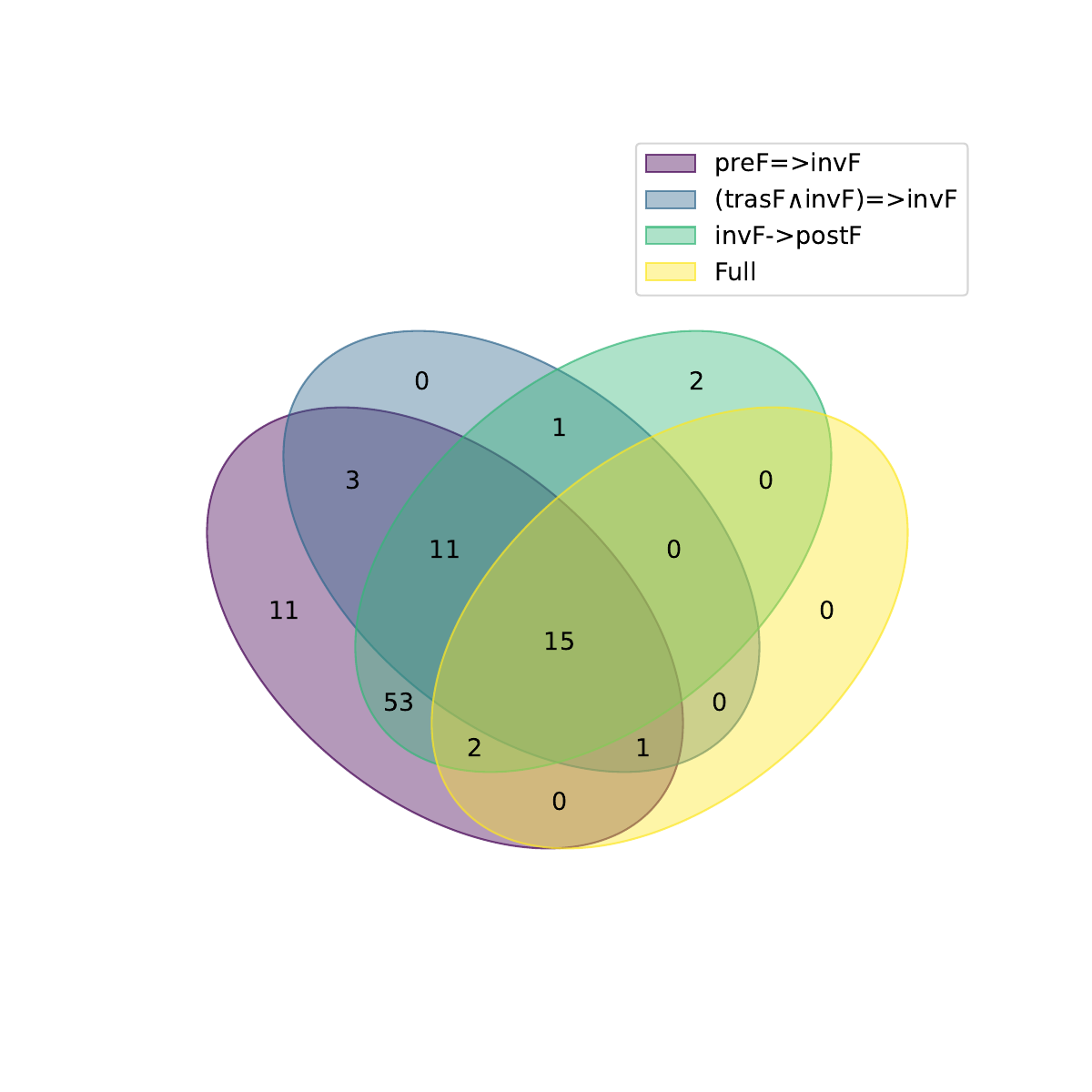}}%
\label{fig:gpt-4-venn}
\hfil
\subfloat[]{\includegraphics[width=2.5in]{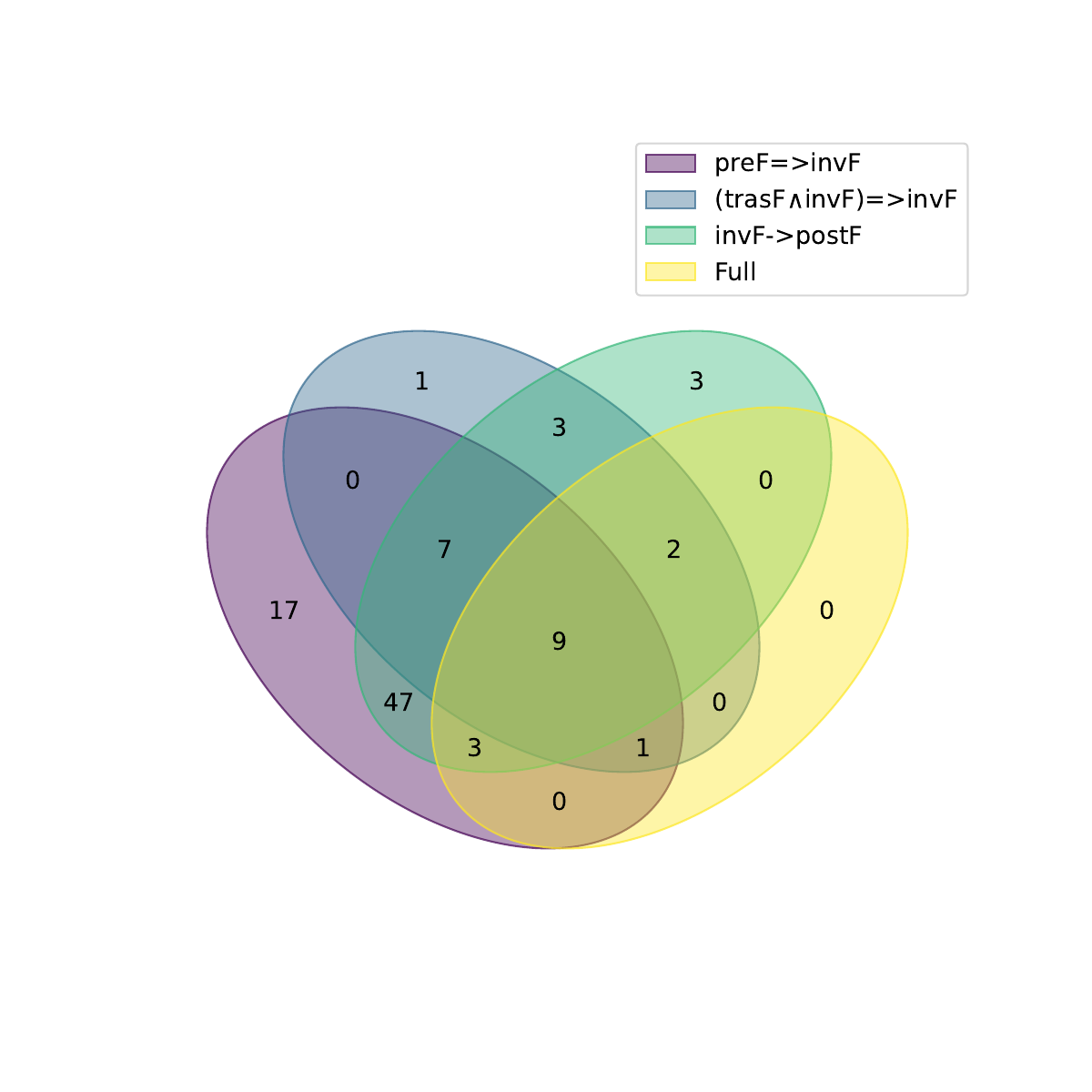}}%
\label{fig:mistral-venn}
\caption{Comparative numbers of sub-problems where LLMs generated correct invariants. (a) GPT-4 generated invariants (b) Mistral generated invariants}
\label{fig:venn}
\end{figure}

Figure \ref{fig:partial-inv-res} provides a summarized figure of the findings. 
Across both the LLMs, we observed that simpler conditions such as equation \ref{c2} and equation \ref{c4} had higher success rates than the more complex condition (\ie equation \ref{c3}). 
For example, for conditions \ref{c2} and \ref{c4}, GPT-4o's success rate is 45\% and 40\%, respectively, while for condition \ref{c3}, it is merely 15\%. 

When combined with the $\mathcal{P}_{full}$, the performance is lower than the performance on individual conditions. To understand the results better, we analyzed the subset of problems where LLMs could generate invariants satisfying each of these conditions. 
\Cref{fig:venn} shows the intersection of problems where LLM could generate a successful solution with the partial prompts and full prompt. 

In the case of GPT-4, there are 26 problems, where LLMs generate correct solution for each of the three $\mathcal{P}_{partial}$s. However, only 15 of those problems were solved with the $\mathcal{P}_{full}$. For Mistral, 16 problems had solutions satisfying all three prompts separately, but when combined, only 9 problems were solved. 

We analyzed the generated invariants with $\mathcal{P}_{full}$ and $\mathcal{P}_{partial}$s to understand why LLM failed to produce satisfying $\mathcal{I}_{final}$, even though they found satisfying $\mathcal{I}_{conditional}$ for all three conditions.

We conjecture that integrating logical expressions (\ie $\mathcal{I}_{conditional}$) poses computational challenges, potentially being undecidable ~\cite{si2018learning}. The generation of valid $\mathcal{I}_{conditional}$s was constrained to a fixed quantity, with no assurance that a valid $\mathcal{I}_{final}$ could emerge from any combination of them. In addition, we anecdotally tested ten (10) problems with manually constructed $\mathcal{P}_{full}$ containing at least one ground truth invariant. Analyzing the generated $\mathcal{I}_{final}$s ($k=50$), we found that none of the generated invariants matched the input ground truth invariant, suggesting that LLMs struggle to infer correct invariants even when provided with strong supporting examples.

\addtocounter{o}{1}\lessons{\theo. Dividing the loop invariant synthesis problem into subproblems and then combining their solutions results in a lower success rate, reaching a maximum of only 8\% in our experimental setup. In contrast, the earlier approach utilizing instructions yielded a higher success rate of 49\%.}

\subsection{RQ1.3 Can few-shot prompting with similar examples impact the invariant synthesis?}

\subsubsection{Motivation}

In RQ1.1 and RQ1.2, we examined how domain knowledge influences the generation of loop invariants by LLMs, achieving a success rate of up to 49\%. While this result demonstrates potential, it is done in a zero-shot setup. Prior research, such as \cite{brown2020language, yoshida2024impact, misu2024towards}, has shown that LLMs are particularly adept at learning from examples, i.e., in a few-shot setting. 


\begin{table*}[t]
    \caption{Comparison of Similarity Measure}
    \centering
    \footnotesize{
    \begin{tabular}{c|c|c}
        \textbf{Query Problem} & \textbf{Example Picked by $\mathcal{S}_{semantic}$} & \textbf{Example Picked by $\mathcal{S}_{syntactic}$} \bigstrut \\
        \hlineB{2}

         \includegraphics[width=0.3\linewidth]{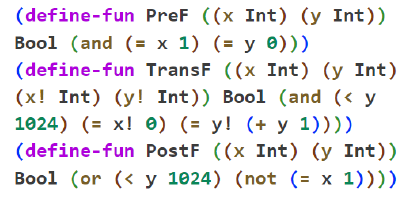}
         &          \includegraphics[width=0.3\linewidth]{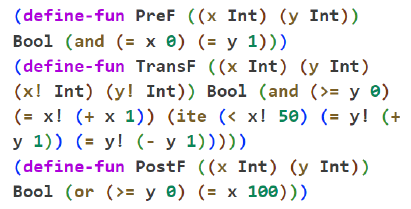} 
          &        \includegraphics[width=0.3\linewidth]{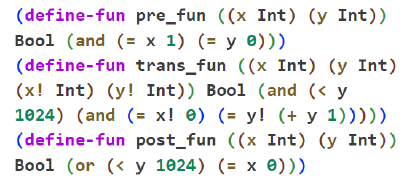} \bigstrut \\
          \hlineB{2}
    \end{tabular}
    }
    \label{tab: similarity}
\end{table*}
\begin{figure} [t]
    \centering
    \includegraphics[width=\linewidth]{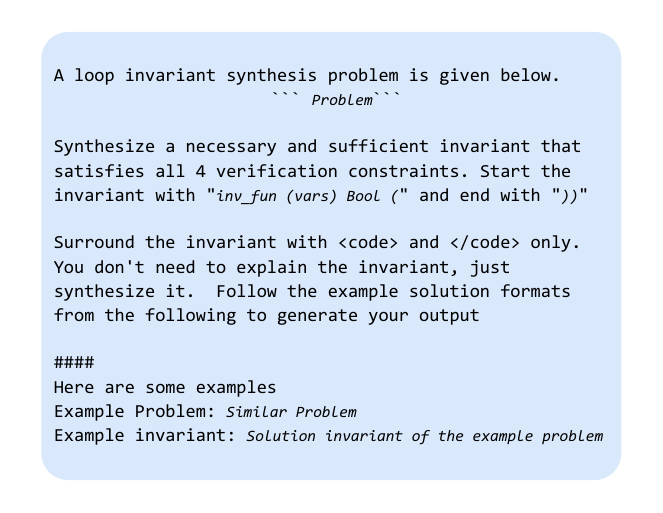}
     \caption{Prompt to generate invariants using few-shot examples} 
    \label{fig:few-shot-template}
\end{figure}
\subsubsection{Approach} For a given problem, we provided an LLM with examples of similar problems and their solutions from our dataset. To find similar examples, we utilized two algorithms. The first algorithm is semantic-based, referred to as $\mathcal{S}_{semantic}$. The second algorithm is syntactic-based, referred to as $\mathcal{S}_{syntactic}$. We describe the algorithms below. 

\paragraph{$\mathcal{S}_{semantic}$ : } We check the semantic similarity between the input problem and other candidate problems in our dataset. Instead of using simply textual similarity, we utilize a tree-based similarity approach to preserve the semantic nature of the problems. We apply the Web STS BERT model\footnote{\url{https://pypi.org/project/semantic-text-similarity/}}~\cite{ahn2022practical} to compute the semantic similarity  ($\mathcal{S}_{semantic}$) between a pair of expressions. 

\paragraph{$\mathcal{S}_{syntactic}$ : }
Computing $\mathcal{S}_{syntactic}$ is more involved than $\mathcal{S}_{semantic}$. Given a problem $\mathcal{P}$ and a candidate $\mathcal{Q}_i$, we parse their precondition (\texttt{PreF}), postcondition (\texttt{PostF}), and transfer function (\texttt{TransF}) into ASTs ~\cite{ragkhitwetsagul2018comparison}, and compute APTED\footnote{\url{https://pypi.org/project/apted/}} similarity between each corresponding pair (\eg, $PreF_{\mathcal{P}}$ vs.\ $PreF_{\mathcal{Q}_i}$). The overall similarity is the mathematical average  
of the three function-level similarities.

\Cref{tab: similarity} illustrates a query problem $\mathcal{P}$ and its top matches under both similarity metrics. $\mathcal{S}_{semantic}$ favors surface-level matches, often selecting examples with identical method names but differing logic. In contrast, $\mathcal{S}_{syntactic}$ captures deeper structural similarity.


We used the prompt in ~\Cref{fig:few-shot-template} for this experiment. We apply the algorithms $\mathcal{S}_{semantic}$ and $\mathcal{S}_{syntactic}$ to retrieve a small set of positive examples, $\mathcal{EX}_{P}$ (with $n=2$) to populate the few-shot prompt. Since the objective of this experiment is to evaluate the performance of the similarity algorithms, we intentionally omit discussion of the rationale behind the types of examples and their quantity, assuming that any such effects are consistent across algorithms. To assess the performance of the LLMs, we evaluate the generated results using sample sizes of $k = 10$, $30$, \& $50$.




\begin{table}[!t]
    \centering
    \caption{Performance comparison of LLM models with instructions.}
    \footnotesize
    { 
       
        \begin{tabular}{c|c|c|c|c|c|c}
        & \multicolumn{6}{c}{\textbf{\% of problems solved in k generated responses}}\bigstrut \\
        \cline{2-7}
        \multirow{2}{*}{\textbf{LLM}} & \multicolumn{3}{c|}{\textbf{Few-shot with $\mathcal{S}_{semantic}$}} & \multicolumn{3}{c}{\textbf{{Few-shot with $\mathcal{S}_{syntactic}$}}}\bigstrut[t] \\
        & $k=10$ & $k=30$ & $k=50$ & $k=10$ & $k=30$ & $k=50$ \bigstrut[b] \\
        \hlineB{2}
        Mistral-large & 29 & 42 & 66 & 53 & 67 & 73  \bigstrut\\ 
        GPT-4o & 43 & 57 & 72 & 61 & 72 & 76 \bigstrut \\
        \hlineB{2}
        \end{tabular}
    }
    \label{tab:fewshot_performance}
\end{table}



\subsubsection{Result} 


Table \ref{tab:fewshot_performance} presents the comparative analysis of the $\mathcal{S}_{semantic}$ and $\mathcal{S}_{syntactic}$ similarity-based selection strategies for few-shot prompting. For the sample size, $k=10$, the $\mathcal{S}_{syntactic}$ yielded a higher success rate (61\% for GPT-4o and 53\% for Mistral-large) compared to the $\mathcal{S}_{semantic}$ (43\% and 31\% respectively). \addtocounter{o}{1}\lessons{\theo. With GPT-4o, for the lower number of sample size ~\ie $k=10$, the performance difference for the algorithms is 18\%, while with a higher sample size $k=50$, the performance difference is as small as 4\%}
With both models, $\mathcal{S}_{syntactic}$ continues to show better performance with the larger batch sizes, $k=30$ and $k=50$. Notably, for $k=50$, $\mathcal{S}_{syntactic}$ achieved the highest success rate of 76\% with GPT-4o, while $\mathcal{S}_{semantic}$ produced a slightly lower success rate of 72\%. These results suggest that as the sample size increases, the difference in performance between the two similarity measures diminishes. 

\addtocounter{o}{1}\lessons{\theo. The syntactic similarity measure yielded up to 76\% success rate for the sample size $k=50$ with few-shot learning.}

\addtocounter{o}{1}\lessons{\theo. Both similarity-based few-shot prompting strategies outperform the instruction-only baseline. Specifically,  $\mathcal{S}_{syntactic}$ achieved a success rate of 76\%, while $\mathcal{S}_{semantic}$  reached to 72\%. In comparison, the instruction-only setup resulted in a lower success rate of 49\%, representing that using relevant examples in few-shot learning helps the model perform better. }



\subsection{RQ1.4 Can the type of the examples impact few-shot prompting for invariant synthesis?}

\subsubsection{Motivation}

The primary philosophy behind in-context learning~\cite{brown2020language} is to show LLM some illustration of how a task is performed. 
In a way, such a prompting strategy is to guide the model on {\em what to do}. 
Few recent works~\cite{hamdan2025much, gao2024customizing} also investigated LLMs' responses when provided with negative examples, guiding the LLMs on {\em what not to do}. 
Inspired by these works, we are motivated to examine the impact of types of few-shot examples in the prompt. 

\subsubsection{Approach}
We constructed few-shot prompts with the following three configurations:
\begin{itemize}[itemsep=0pt, leftmargin=10pt]
\item $\mathcal{EX}_{P}$ (Positive example) : A positive example consists of a problem paired with an invariant that satisfies the specification. 
\item $\mathcal{EX}_{N}$ (Negative example) : A negative example is an invariant synthesis problem, a candidate invariant that fails to satisfy the specification, and the reason for this failure. 
\item $\mathcal{EX}_{Mix}$ (Mixed example) : The Mixed example is composed of both satisfying and unsatisfying invariants of a problem.
\end{itemize}





\Cref{tab:examples} shows examples of  of $\mathcal{EX}_{P}$, $\mathcal{EX}_{N}$, and $\mathcal{EX}_{Mix}$ and how we present them in the prompt.
 We adopted the same experimental setup as RQ1.3, using examples retrieved via the $\mathcal{S}_{syntactic}$, which was better than $\mathcal{S}_{semantic}$ in RQ1.3.

\begin{table*}[!t]
    \centering
    \caption{Different types of Examples based on their verification status for few-shot prompt 
    }
    \footnotesize{
    \begin{tabular}{m{0.1\linewidth}|m{0.7\linewidth}}
        \textbf{Type} & \textbf{Example}  \bigstrut \\ 
        \hlineB{2} \\
         $\mathcal{EX}_{P}$ & \includegraphics[width=\linewidth]{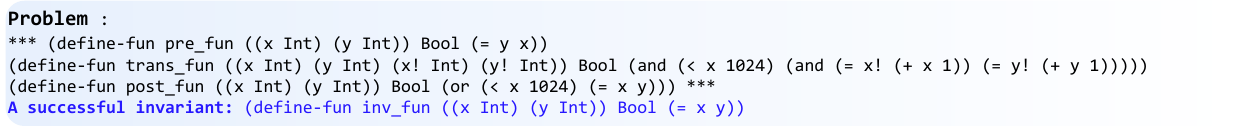}
         \bigstrut \\
         $\mathcal{EX}_{N}$ & \includegraphics[width=\linewidth]{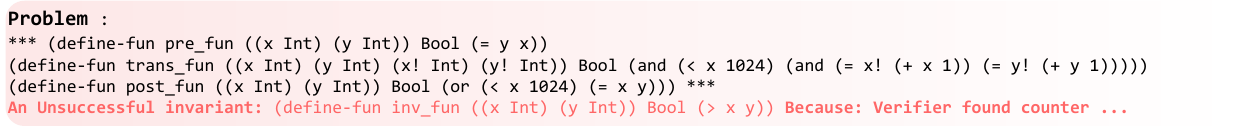} \bigstrut \\
         $\mathcal{EX}_{Mix}$ & \includegraphics[width=\linewidth]{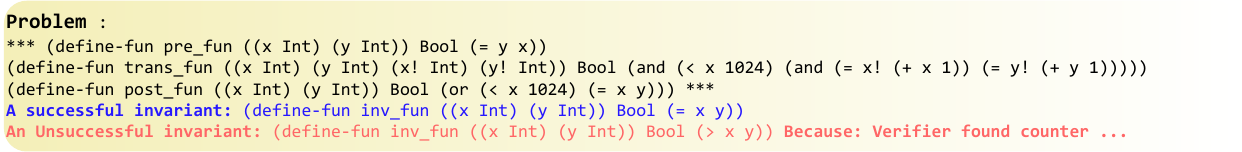} \bigstrut \\
         \hlineB{2}
    \end{tabular}    
    }
    \label{tab:examples}
\end{table*}
\subsubsection{Result}

\begin{table}[!t]
    \centering
    \caption{Study on LLM's Successful Invariant Generation with few-shot prompts}
    \footnotesize {
        \begin{tabular}{c|c|c|c|c}
         & \multicolumn{4}{c}{\textbf{\% of problems solved in k generated responses}}\bigstrut \\
         \cline{2-5}
         \multirow{2}{*}{\textbf{LLM}} & \multirow{2}{*}{\textbf{Zero shot}} & \multicolumn{3}{c}{\textbf{Few-shot}}\bigstrut[t] 
         \\
       &  & $\mathcal{EX}_{P}$ & $\mathcal{EX}_{N}$ & $\mathcal{EX}_{Mix}$
        \bigstrut[b] \\
        \hlineB{2}
        Mistral-large & 21\% & \textbf{73\%} & 23\% & 68\%  \bigstrut\\  
        GPT-4o&  40\% & \textbf{76\%} & 34\% & 71\%  \bigstrut\\ 
         \hlineB{2}
    \end{tabular}
    }   
      
    \label{tab: few-shotComparison}
\end{table}
 
Table \ref{tab: few-shotComparison} shows that both LLMs perform better when given only positive examples in the few-shot prompt. With $\mathcal{EX}_{P}$s, GPT-4o achieves a success rate of 76\%. In contrast, $\mathcal{EX}_{N}$s and  $\mathcal{EX}_{Mix}$ under the same setup yield lower success rates of 34\% and 61\%, respectively. We examined individual cases to understand how LLMs learn from examples. For example  a problem with the ground truth invariant {\texttt{(and (>= x 1) (<= x 101))}}, $\mathcal{EX}_{P}$ provided the invariant - {\texttt{(>= x 1))}} as example in the prompt, and  GPT-4o produced the invariant  {\texttt{(and (>= x 1) (<= x 101))}}, that is exactly same as the ground truth, on the other hand $\mathcal{EX}_{N}$ provided 
{\texttt{(and (<= x 11) (= y (- 10 x)}} as example, and GPT-4o produced {\texttt{(and (<= x 101) (> y (- 101 x)))}} as solution invariant.
This suggests that while negative examples help the model avoid certain incorrect patterns, they do not necessarily guide it toward the correct one due to the vast space of possible incorrect invariants. 

Furthermore, our results show that GPT-4o performs better in a zero-shot setting (40\%) than when prompted with $\mathcal{EX}_{N}$ examples (34\%). This finding further underscores the advantage of positive examples in effectively guiding the synthesis of correct invariants.



\addtocounter{o}{1}\lessons{\theo. The few-shot (positive-only) approach achieved the highest success rate of 76\%, outperforming other example types discussed.}






\subsection{RQ1.5 Can we combine domain knowledge with few-shot examples to synthesize more accurate loop invariant? }

\subsubsection{Motivation}
Our previous experiments offer insights on the usefulness of providing instructions with domain knowledge and positive examples in few-shot settings. It is thus necessary to determine whether we can achieve even better performance by combining both strategies.

\begin{table}[!t]
    
    \caption{Performance comparison of LLMs with Combined Few-shot and Instruction-based prompts}
    \centering
    \footnotesize{
    \begin{tabular}{p{.3\linewidth}|p{.15\linewidth}|p{.15\linewidth}|p{.15\linewidth}}
        & \multicolumn{3}{c}{\textbf{\% of problems solved in k generated responses}} \bigstrut \\
        \textbf{Prompt Setup} & $k=10$ & $k=30$ & $k=50$ \bigstrut[b] \\
        \hlineB{2}
        Instruction only & 27\% & 41\% & \textbf{49\%} \bigstrut \\   
        Few-shot only & \textbf{61\%} & \textbf{72\%} & \textbf{76\%} \bigstrut\\ 
        Integrated Few-Shot with Instruction & 57\% & \textbf{73\%} & \textbf{78\%} \bigstrut\\         
        \hlineB{2}
    \end{tabular}  
    }
    \label{tab:combined}
\end{table}

\subsubsection{Approach}

\begin{figure}[!ht]
    \centering
    \includegraphics[width=\linewidth]{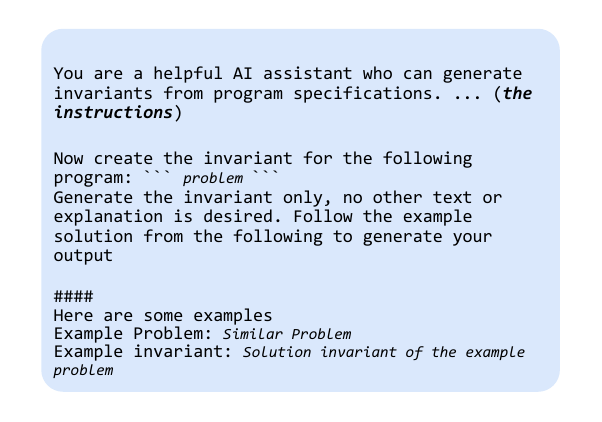}
    \caption{Prompt Template to generate invariants using the combination of instructions and few-shots with LLM}
    \label{fig:prompt_combined_gi_few}
\end{figure}

The experiment is designed by updating the prompts, appending the few-shot examples to the set of instructions within the prompt. The prompt structure used in this setup is illustrated in ~\Cref{fig:prompt_combined_gi_few}, which first presents the general instructions (as described in RQ1.1), followed by the target problem, and finally the selected few-shot examples. In this phase, we focused the experiment exclusively on GPT-4o, as it demonstrated the best performance across RQ1.1 to RQ1.4.     


\subsubsection{Result}

The experiment combining instruction-based prompts with few-shot examples is summarized in Table \ref{tab:combined}. Excluding the cases where LLM's token limit was exceeded, the success rates of the generated invariants were 57\%, 73\%, and 78\% for $k = 10, 30, and 50$, respectively. The integrated approach yielded performance comparable to using only the few-shot positive examples, with a maximum improvement of just 2\% across all sample sizes. However, it is notable that 53\% of the prompts in the integrated setup exceeded the 8k token limit, resulting in failed requests.

\addtocounter{o}{1}\lessons{\theo. With an integrated approach combining instructions and few-shot prompting, GPT-4o achieved a success rate of up to 78\% for $k=50$, the highest across all our experiments.}

%% file: body/RQ2.tex
\section{Loop Invariant Repair by LLMs (RQ2)}

\subsection{RQ2.1 Can feedback to an LLM help it repair an invariant which it generated incorrectly? }

\subsubsection{Motivation}
Upon observing a large number of failed invariants in RQ1, we investigated the underlying cause of the error by analyzing the error messages generated from the verifier.

The feedback from the verifier highlights the specific cause of the error and offers a detailed message describing the issue. For example, one example feedback is: {\texttt{ \{cause: FAIL - internal exceptions (Unsatisfied) \}, \{details: the verifier failed to satisfy the
condition ( => (PreF n i k j) (InvF n i k j)))) \} }}, indicating the error of not satisfying the verification condition $PreF => InvF$. Such error details inspired us to explore the possibility of correcting the invariant. 

\begin{figure}
    \centering
    \includegraphics[width=\linewidth]{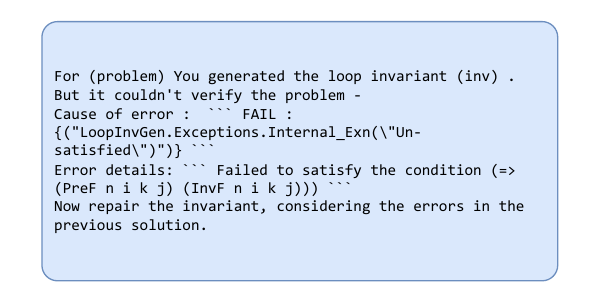}
    \caption{Prompt Template to repair incorrect invariant }
    \label{fig:template_rep}
\end{figure}


\subsubsection{Approach}

We instructed the LLM to generate invariants for all problems in our benchmark. Each generated invariant was subsequently tested for correctness. If an invariant was found to be incorrect, we recorded the corresponding feedback from the verifier. 
For instance, in the example provided in the figure \ref{fig:template_rep}, the verifier could not satisfy the first rule (\ref{c2}). We prepared the invariant fixing prompt (~\Cref{fig:template_rep}) capitalizing on that feedback.
For every problem, we selected the first two incorrect invariants and asked the LLM to repair them using the verifier's feedback, which includes an error cause (in form of title) and a detailed error message. We incorporated this feedback into the prompt. 

\subsubsection{Result}
\begin{table}
    \caption{LLMs Performance in Invariant Repair Using Errors}
    \centering
    \footnotesize{
    \begin{tabular} {c|c|c|c}
    
         LLM Model&  \# Problems &   \# Attempted invariants & \%  Successful repair \bigstrut \\
         \hlineB{2}
         GPT-4o&  210&   420& 6\% \bigstrut \\ 
         Mistral-Large&  210&   420& 4\% \bigstrut \\ 
         \hlineB{2}
    \end{tabular}   
    }
    \label{tab:llmRepairInv}
\end{table}



Table \ref{tab:llmRepairInv} presents the results of our study evaluating the effectiveness of LLMs in repairing incorrect invariants. Out of 420 generated incorrect invariants (the first 2 incorrect invariants for each of the 210 problems), GPT-4 successfully repaired only 6\%, while Mistral-large achieved a slightly lower success rate of 4\%. 


 Evaluating the successful repairing cases, we observed that, in some cases, the invariant is fixed by removing the unnecessary words \ie \textit{code, scheme, lisp etc.} that caused an expression error. For example, for the following repair case (Listing \ref{lst:ex1}), an additional keyword \textit{code} was added in the initially generated invariant, LLM fixed the error by recognizing "Expression error" from the feedback. 

\begin{lstlisting}[caption={Repair attempt example 1},label={lst:ex1}]
Failed inv: code ((...) Bool (and (= conf_0 8) (>= y 0)))
Repaired inv: ((...) Bool (and (= conf_0 8) (>= y 0)))
\end{lstlisting}

 There were several cases where LLM didn't remove the additional word. For example (Listing \ref{lst:ex2} ), in the following case, LLM updated the  expression instead of removing the expression error


\begin{lstlisting}[caption={Repair attempt example 2},label={lst:ex2}]
Failed inv: code Bool (and (= conf_0 3) (= c_1 0)(> n_0 0))
Repaired inv: code Bool   (and (= conf_0 3) (= c 0) (> n 0)))
\end{lstlisting}

Another successful error-fixing pattern was removing unnecessary conditions from the previously generated invariant. For example, in Listing \ref{lst:ex3}, the conditions $j > 0$ and $k > 0$ were removed from the initially generated invariant.  

\begin{lstlisting}[caption={Repair attempt example 3},label={lst:ex3}]
Failed inv: code Bool (and (<= 0 i)(<= i (* j k))(> j 0)(> k 0))
Repaired inv: code Bool (and (<= 0 i) (<= i (* j k)))
\end{lstlisting}

Also, we noticed that, in some cases (Listing \ref{lst:ex4}), LLM changed the complete expression to fix the error. 

\begin{lstlisting}[caption={Repair attempt example 4},label={lst:ex4}]
Failed inv: code Bool (and (= c c_2) (>= c_2 0) (<= c_2 4))
Repaired inv: code Bool (and (>= c 0) (<= c 4))
\end{lstlisting}

These results demonstrate that, even with verifier feedback, both models struggled to consistently repair incorrect invariants. 

\addtocounter{o}{1}\lessons{\theo. Several patterns in fixing the invariants to a successful one were identified, but LLM was inconsistent in following the patterns. Overall, with the error information,  GPT-4o achieved a repair success rate of only 6\%, slightly outperforming Mistral-large, which reached 4\%.}





\subsection{RQ2.2 Does providing information about failure root causes and logs help an LLM repair an incorrect invariant?}

\subsubsection{Motivation}


In Z3, satisfiability is verified by systematically exploring the domain values of all variables, and the solver logs specific combinations of values for which the verification fails. As shown in the error log in the following Listing (~\ref{lst:log_model}), the key \texttt{condition} specifies the violated verification condition, in this case, \texttt{InvF} $\Rightarrow$ \texttt{PostF}, and the key \texttt{model} provides a concrete counterexample. For instance, the values \texttt{x = 1}, \texttt{y = 1024}, \texttt{x! = 0}, and \texttt{y! = 0} led to the failure of the implication. These logs prompted us to explore whether incorporating such failure-specific diagnostic information can enhance the LLM's ability to repair.

\begin{lstlisting}[caption={Error log},label={lst:log_model}]
condition: (=> (inv_fun x y) (post_fun x y)))
model: (define-fun x () Int 1) (define-fun y () Int 1024)
    (define-fun x! () Int 0) (define-fun y! () Int 0)
\end{lstlisting}


\subsubsection{Approach}


\begin{figure}
    \centering
    \includegraphics[width=\linewidth]{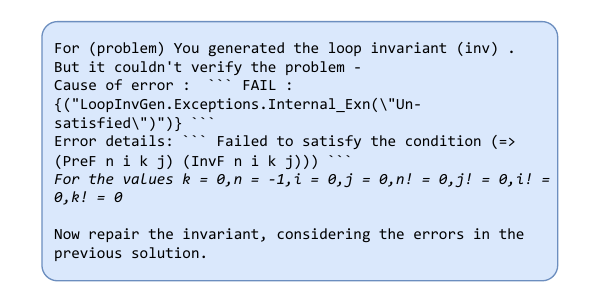}
    \caption{Prompt to repair invariant using counterexamples}
    \label{fig:template_rep2}
\end{figure}

We updated the prompt used in \textbf{RQ2.1} by adopting the model information to explicitly specify the unsatisfactory verification conditions and the corresponding counterexample that contributed to the verification failure. To ensure consistency, we maintained the same experimental setup as in \textbf{RQ2.1}. 


\subsubsection{Result}

\begin{table}
    \caption{LLMs Performance in Invariant Repair Using Error Values}
    \centering
    \begin{tabular} {c|c|c|c}
    
         LLM Model&  \# Problems &   \# Attempted invariants & \%  Successful repair \bigstrut \\
         \hlineB{2}
         GPT-4o&  210&   420& 16\% \bigstrut \\ 
         Mistral-Large&  210&   420& 7\% \bigstrut \\ 
         \hlineB{2}
    \end{tabular}   
    \label{tab:llmRepairInvValues}
\end{table}

Table \ref{tab:llmRepairInvValues} presents the results of our experiment. When provided with detailed error information, GPT-4 successfully repaired 16\% of the incorrect invariants, while Mistral-large achieved a success rate of 7\%. Both models demonstrated improvement compared to the baseline setup in \textbf{RQ2.1}, where they exhibited significantly lower success rates. These results indicate that explicit details about unsatisfied conditions and the variable values causing verification failures enable LLMs to make more informed corrections. However, despite this improvement, the overall success rates remain relatively low, suggesting that LLMs still struggle to fully understand and effectively utilize verifier feedback for invariant repair. 



\addtocounter{o}{1}\lessons{\theo. Although achieving a relatively low success rate of up to 16\%, LLMs perform marginally better at repairing incorrect invariants when provided with the exact variable values that caused the verification errors}


\begin{figure}[t]
    \centering
    \includegraphics[width=1\linewidth]{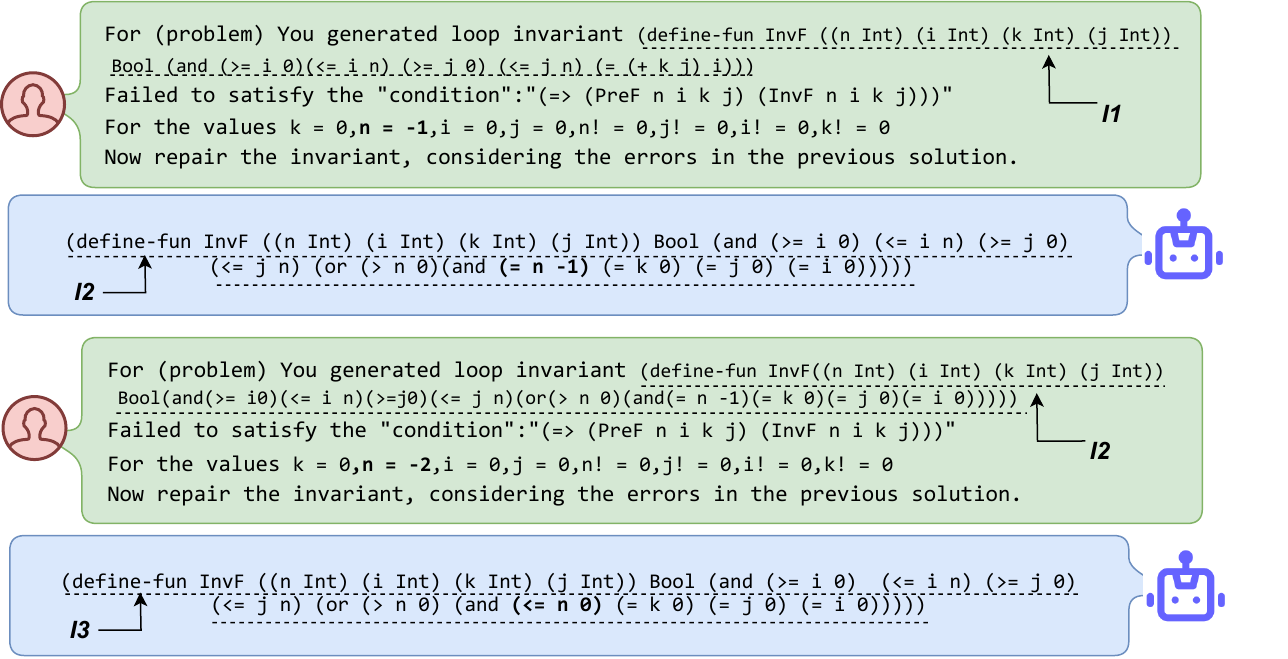}
    \caption{Invariant repairing with LLM}
    \label{fig: repair_steps}
\end{figure}


To understand the LLM's invariant repair process, we closely examined a specific case, illustrated in Figure \ref{fig: repair_steps}. Our observations reveal that when the LLM is provided with detailed information for repairing an invariant, it attempts to address the specific failure case mentioned in the prompt during its first revision. For example, in the case shown in Figure \ref{fig: repair_steps}, the invariant (I1) failed for the case \textit{n = -1}. The LLM responded by repairing the invariant and producing a new version (I2). While I2 successfully passed the case \textit{n = -1}, it failed for a new case, \textit{n = -2}. A subsequent repair request led the LLM to generate another invariant (I3), which satisfied \textit{n = -2} but failed for a different case. This iterative process continued, with the LLM eventually recreating earlier invariants, such as I1, and repairing them again into I2, creating a loop of repeated, incorrect invariants. This behavior suggests that while LLMs can make localized fixes based on specific failures, they often lack a global understanding of the invariant space, highlighting the need for strategies that promote generalization rather than reactive generation.

\addtocounter{o}{1}\lessons{\theo. With the exclusive information about the error, LLMs tend to make case-specific fixes that lack generalization.}






%% file: body/5.discussion.tex
\section{Threats to Validity}
The validity of the study can be threatened by several factors. First, the validity of the study can be compromised by the quality of the problem scenario. In this study, we examined the solvable problems and the invariants that are feasible to generate. The quality of the generated invariants can be influenced by the quality of the prompt, the quality of the LLM, and the quality of the few-shot examples. Second, for the open-source LLMs we examined, the processing unit we used may have adversely affected the performance of the LLMs. We observed that open-source LLMs are way better with their larger models as well as with their interactive chat UIs. Third, the validity of the study can be threatened by the quality of the Z3 solver. We didn't question the quality of the Z3 solver that comes with the solution prepared by \cite{padhi2016data}. Fourth, the LLM-generated responses are natural texts; in many cases, those are not syntactically correct or contain imbalanced parentheses. We have to manually correct those before feeding them to the Z3 solver. This manual correction can be a threat to the validity of the study. Further, the non-deterministic generative nature of the LLMs keeps the door open to question the reproducibility of the same set of outputs.    



%% file: body/7.related.tex
\section{Related Works on Loop Invariant Synthesis}
\noindent\textbf{Traditional Approaches.}
Among the earliest, Hoare et al. \cite{hoare1969axiomatic} introduced deductive methods based on Hoare logic for loop invariant synthesis. These methods required manual crafting of invariants or deriving them using fixed templates. Automated approaches, such as abstract interpretation by Michael \cite{karr1976affine}, and Cousot et al. \cite{cousot1978automatic}, computed invariants over abstract domains. Col{\'o}n et al. \cite{colon2003linear} presented a linear invariant generation technique using symbolic executors. 
Constraint-based methods by Gulwani et al. \cite{gulwani2008program} synthesized invariants by solving systems of equations derived from program semantics. Besides these, Satisfiability Modulo Theory (SMT) based approaches gained the focus of the community on a larger scale  \cite{de2008z3, flanagan2001houdini, flanagan2002predicate, corr18padhi, lahiri2007predicate, gulwani2009constraint, fedyukovich2018accelerating}. The authors of \cite{echenim2019ilinva, ernst2007daikon, le2019sling},  discussed the impact of dynamic analysis along with symbolic execution for invariant study. \cite{jhala2006practical, mcmillan2010lazy} implemented the Craig Interpolation to verify the invariants. Constraint solving \cite{gupta2009tests}, syntax-guided-synthesis \cite{barrett2011cvc4} \cite{barbosa2022cvc5}, counter-example guided abstraction refinement \cite{clarke2003counterexample, zhu2015learning} were the other significant approaches in traditional approach. While these methods were effective for very limited classes of programs such as linear arithmetic or simple program structures, they often required significant manual effort and struggled to handle complex programs with non-linear arithmetic or intricate memory manipulations.

\noindent\textbf{Data-Driven and Machine Learning Approaches.}
The advent of machine learning has led to data-driven methods for loop invariant synthesis. Sharma et al. \cite{sharma2013data} introduced the use of decision trees and support vector machines to infer invariants from execution traces. Later, \cite{sharma2016invariant, padhi2016data} extended the decision tree to learn from the counterexamples to generate invariants. Active learning on top of the decision tree to utilize the feedback from the verifier and modify the invariants was used by \cite{garg2016learning, ezudheen2018horn}.  Deep learning models, such as recurrent neural networks (RNNs) and graph neural networks (GNNs), have been applied to learn invariants from program representations like abstract syntax trees (ASTs) or control-flow graphs \cite{si2018learning, yu2023loop}. Yao et al. \cite{yao2020learning} used Gated Continuous Logic Networks (GCLN) to learn the loop invariants. Code2Inv \cite{si2020code2inv} proposed another deep-learning-based framework to generate loop invariants. These methods leverage large datasets of annotated programs to train models, enabling them to predict invariants for new programs. However, data-driven approaches primarily depended on extensive datasets and were limited by learning programming scenarios. Additionally, the interpretability of machine learning-generated invariants is often limited by difficulty in understanding the reasoning behind their predictions.

\noindent\textbf{LLM-Based Approaches}. The success of LLMs like GPT-3 \cite{brown2020language}, Codex \cite{chen2021evaluating}, and PaLM \cite{chowdhery2022palm}  in code generation and understanding has motivated the research paradigm of invariant generation. Pei et al. examined the ability of fine-tuned LLM in generating loop invariants \cite{pei2023can}. Chakraborty et al. presented an algorithm to rank the LLM-generated invariants to identify the correct invariant from a set of invariants faster \cite{chakraborty2023ranking}. Kamath et al. presented a generative model to generate inductive loop invariants for basic C problems \cite{kamath2023finding}. Lemur \cite{wu2023lemur} introduced a neuro-symbolic approach that combines LLMs with static analysis, leveraging their implicit inductive biases to generate candidate invariants for verification tasks. Rakib et al. used LLM with Chain of Thought (COT) to generate invariants for Dafny problems \cite{misu2024towards}. ACInv \cite{liu2024enhancing} integrated a static analysis with an LLM to generate a loop invariant for a complex program. Pirzada et al. \cite{pirzada2024llm} used LLMs to generate loop invariants by transforming loops into CFG nodes, enabling verification without loop unrolling. However, the correctness of the generated invariants in these papers is either not guaranteed, or the requirement on many candidate invariants was very high. Another shortcoming of these approaches is dependency on a programming language analyzer, which makes them less generalizable.

%% file: body/8.conclusion.tex
\section{Conclusion}

We have studied the extended nature of LLM models in the context of inductive loop invariant synthesis for low program specifications. We show that LLMs can be effective for the loop invariant synthesis task, with GPT-4o demonstrating the highest success rate and consistency. Mistral-large also showed closer performance. Few-shot prompting was significantly better than an instruction-only setup, especially when examples were selected based on syntactic similarity. Though the best results were achieved by combining instructions with few-shot examples, this strategy is limited by input token constraints. We attempted to decompose the synthesis task into subproblems, but it was less effective. While LLMs often solved individual sub-conditions, they struggled to integrate the solutions into a valid invariant, suggesting difficulty in combining logical expressions coherently.  In the invariant repairing task, LLMs showed limited effectiveness, even with the detailed error information. This observation highlights the need for more robust prompting/tuning strategies. Future work could explore ways to overcome this limitation to further enhance performance.



%% file: main.bbl
\begin{thebibliography}{10}
\providecommand{\url}[1]{#1}
\csname url@samestyle\endcsname
\providecommand{\newblock}{\relax}
\providecommand{\bibinfo}[2]{#2}
\providecommand{\BIBentrySTDinterwordspacing}{\spaceskip=0pt\relax}
\providecommand{\BIBentryALTinterwordstretchfactor}{4}
\providecommand{\BIBentryALTinterwordspacing}{\spaceskip=\fontdimen2\font plus
\BIBentryALTinterwordstretchfactor\fontdimen3\font minus \fontdimen4\font\relax}
\providecommand{\BIBforeignlanguage}[2]{{%
\expandafter\ifx\csname l@#1\endcsname\relax
\typeout{** WARNING: IEEEtran.bst: No hyphenation pattern has been}%
\typeout{** loaded for the language `#1'. Using the pattern for}%
\typeout{** the default language instead.}%
\else
\language=\csname l@#1\endcsname
\fi
#2}}
\providecommand{\BIBdecl}{\relax}
\BIBdecl

\bibitem{karr1976affine}
M.~Karr, ``Affine relationships among variables of a program,'' \emph{Acta Informatica}, vol.~6, no.~2, pp. 133--151, 1976.

\bibitem{furia2014loop}
C.~A. Furia, B.~Meyer, and S.~Velder, ``Loop invariants: Analysis, classification, and examples,'' \emph{ACM Computing Surveys (CSUR)}, vol.~46, no.~3, pp. 1--51, 2014.

\bibitem{si2018learning}
X.~Si, H.~Dai, M.~Raghothaman, M.~Naik, and L.~Song, ``Learning loop invariants for program verification,'' \emph{Advances in Neural Information Processing Systems}, vol.~31, 2018.

\bibitem{colon2003linear}
M.~A. Col{\'o}n, S.~Sankaranarayanan, and H.~B. Sipma, ``Linear invariant generation using non-linear constraint solving,'' in \emph{Computer Aided Verification: 15th International Conference, CAV 2003, Boulder, CO, USA, July 8-12, 2003. Proceedings 15}.\hskip 1em plus 0.5em minus 0.4em\relax Springer, 2003, pp. 420--432.

\bibitem{chakraborty2023ranking}
S.~Chakraborty, S.~K. Lahiri, S.~Fakhoury, M.~Musuvathi, A.~Lal, A.~Rastogi, A.~Senthilnathan, R.~Sharma, and N.~Swamy, ``Ranking llm-generated loop invariants for program verification,'' \emph{arXiv preprint arXiv:2310.09342}, 2023.

\bibitem{hrushovski2023strongest}
E.~Hrushovski, J.~Ouaknine, A.~Pouly, and J.~Worrell, ``On strongest algebraic program invariants,'' \emph{Journal of the ACM}, vol.~70, no.~5, pp. 1--22, 2023.

\bibitem{muller2004computing}
M.~M{\"u}ller-Olm and H.~Seidl, ``Computing polynomial program invariants,'' \emph{Information Processing Letters}, vol.~91, no.~5, pp. 233--244, 2004.

\bibitem{flanagan2001houdini}
C.~Flanagan and K.~R.~M. Leino, ``Houdini, an annotation assistant for esc/java,'' in \emph{International Symposium of Formal Methods Europe}.\hskip 1em plus 0.5em minus 0.4em\relax Springer, 2001, pp. 500--517.

\bibitem{corr18padhi}
\BIBentryALTinterwordspacing
S.~Padhi, R.~Sharma, and T.~Millstein, ``Loopinvgen: A loop invariant generator based on precondition inference,'' \emph{CoRR}, vol. abs/1707.02029, 2018. [Online]. Available: \url{http://arxiv.org/abs/1707.02029}
\BIBentrySTDinterwordspacing

\bibitem{lahiri2007predicate}
S.~K. Lahiri and R.~E. Bryant, ``Predicate abstraction with indexed predicates,'' \emph{ACM Transactions on Computational Logic (TOCL)}, vol.~9, no.~1, pp. 4--es, 2007.

\bibitem{padhi2016data}
S.~Padhi, R.~Sharma, and T.~Millstein, ``Data-driven precondition inference with learned features,'' \emph{ACM SIGPLAN Notices}, vol.~51, no.~6, pp. 42--56, 2016.

\bibitem{kamath2023finding}
A.~Kamath, A.~Senthilnathan, S.~Chakraborty, P.~Deligiannis, S.~K. Lahiri, A.~Lal, A.~Rastogi, S.~Roy, and R.~Sharma, ``Finding inductive loop invariants using large language models,'' \emph{arXiv preprint arXiv:2311.07948}, 2023.

\bibitem{misu2024towards}
M.~R.~H. Misu, C.~V. Lopes, I.~Ma, and J.~Noble, ``Towards ai-assisted synthesis of verified dafny methods,'' \emph{Proceedings of the ACM on Software Engineering}, vol.~1, no. FSE, pp. 812--835, 2024.

\bibitem{zeng2023evaluating}
Z.~Zeng, J.~Yu, T.~Gao, Y.~Meng, T.~Goyal, and D.~Chen, ``Evaluating large language models at evaluating instruction following,'' \emph{arXiv preprint arXiv:2310.07641}, 2023.

\bibitem{wu2024divide}
Z.~Wu, H.~Bai, A.~Zhang, J.~Gu, V.~Vydiswaran, N.~Jaitly, and Y.~Zhang, ``Divide-or-conquer? which part should you distill your llm?'' \emph{arXiv preprint arXiv:2402.15000}, 2024.

\bibitem{reynolds2021prompt}
L.~Reynolds and K.~McDonell, ``Prompt programming for large language models: Beyond the few-shot paradigm,'' in \emph{Extended abstracts of the 2021 CHI conference on human factors in computing systems}, 2021, pp. 1--7.

\bibitem{de2008z3}
L.~De~Moura and N.~Bj{\o}rner, ``Z3: An efficient smt solver,'' in \emph{International conference on Tools and Algorithms for the Construction and Analysis of Systems}.\hskip 1em plus 0.5em minus 0.4em\relax Springer, 2008, pp. 337--340.

\bibitem{hoare1969axiomatic}
C.~A.~R. Hoare, ``An axiomatic basis for computer programming,'' \emph{Communications of the ACM}, vol.~12, no.~10, pp. 576--580, 1969.

\bibitem{alur2013syntax}
R.~Alur, R.~Bodik, G.~Juniwal, M.~M. Martin, M.~Raghothaman, S.~A. Seshia, R.~Singh, A.~Solar-Lezama, E.~Torlak, and A.~Udupa, \emph{Syntax-guided synthesis}.\hskip 1em plus 0.5em minus 0.4em\relax IEEE, 2013.

\bibitem{uddin2022empirical}
G.~Uddin, Y.-G. Gu{\'e}h{\'e}nuc, F.~Khomh, and C.~K. Roy, ``An empirical study of the effectiveness of an ensemble of stand-alone sentiment detection tools for software engineering datasets,'' \emph{ACM Transactions on Software Engineering and Methodology (TOSEM)}, vol.~31, no.~3, pp. 1--38, 2022.

\bibitem{meta2024llama3}
M.~AI, ``Meta llama 3.1,'' \url{https://ai.meta.com/blog/meta-llama-3-1/}, 2024, accessed: 2024-07-29.

\bibitem{mistral_large}
\BIBentryALTinterwordspacing
------, ``Mistral large,'' 2024, accessed: 2025-03-08. [Online]. Available: \url{https://mistral.ai/models}
\BIBentrySTDinterwordspacing

\bibitem{achiam2023gpt}
J.~Achiam, S.~Adler, S.~Agarwal, L.~Ahmad, I.~Akkaya, F.~L. Aleman, D.~Almeida, J.~Altenschmidt, S.~Altman, S.~Anadkat \emph{et~al.}, ``Gpt-4 technical report,'' \emph{arXiv preprint arXiv:2303.08774}, 2023.

\bibitem{chang2024survey}
Y.~Chang, X.~Wang, J.~Wang, Y.~Wu, L.~Yang, K.~Zhu, H.~Chen, X.~Yi, C.~Wang, Y.~Wang \emph{et~al.}, ``A survey on evaluation of large language models,'' \emph{ACM transactions on intelligent systems and technology}, vol.~15, no.~3, pp. 1--45, 2024.

\bibitem{wu2023lemur}
H.~Wu, C.~Barrett, and N.~Narodytska, ``Lemur: Integrating large language models in automated program verification,'' \emph{arXiv preprint arXiv:2310.04870}, 2023.

\bibitem{brown2020language}
T.~Brown, B.~Mann, N.~Ryder, M.~Subbiah, J.~D. Kaplan, P.~Dhariwal, A.~Neelakantan, P.~Shyam, G.~Sastry, A.~Askell \emph{et~al.}, ``Language models are few-shot learners,'' \emph{Advances in neural information processing systems}, vol.~33, pp. 1877--1901, 2020.

\bibitem{yoshida2024impact}
L.~Yoshida, ``The impact of example selection in few-shot prompting on automated essay scoring using gpt models,'' in \emph{International Conference on Artificial Intelligence in Education}.\hskip 1em plus 0.5em minus 0.4em\relax Springer, 2024, pp. 61--73.

\bibitem{ahn2022practical}
S.~Ahn, S.~Ahn, H.~Koo, and Y.~Paek, ``Practical binary code similarity detection with bert-based transferable similarity learning,'' in \emph{Proceedings of the 38th Annual Computer Security Applications Conference}, 2022, pp. 361--374.

\bibitem{ragkhitwetsagul2018comparison}
C.~Ragkhitwetsagul, J.~Krinke, and D.~Clark, ``A comparison of code similarity analysers,'' \emph{Empirical Software Engineering}, vol.~23, pp. 2464--2519, 2018.

\bibitem{hamdan2025much}
S.~Hamdan and D.~Yuret, ``How much do llms learn from negative examples?'' \emph{arXiv preprint arXiv:2503.14391}, 2025.

\bibitem{gao2024customizing}
X.~Gao and K.~Das, ``Customizing language model responses with contrastive in-context learning,'' in \emph{Proceedings of the AAAI Conference on Artificial Intelligence}, vol.~38, no.~16, 2024, pp. 18\,039--18\,046.

\bibitem{cousot1978automatic}
P.~Cousot and N.~Halbwachs, ``Automatic discovery of linear restraints among variables of a program,'' in \emph{Proceedings of the 5th ACM SIGACT-SIGPLAN symposium on Principles of programming languages}, 1978, pp. 84--96.

\bibitem{gulwani2008program}
S.~Gulwani, S.~Srivastava, and R.~Venkatesan, ``Program analysis as constraint solving,'' in \emph{Proceedings of the 29th ACM SIGPLAN Conference on Programming Language Design and Implementation}, 2008, pp. 281--292.

\bibitem{flanagan2002predicate}
C.~Flanagan and S.~Qadeer, ``Predicate abstraction for software verification,'' in \emph{Proceedings of the 29th ACM SIGPLAN-SIGACT symposium on Principles of programming languages}, 2002, pp. 191--202.

\bibitem{gulwani2009constraint}
S.~Gulwani, S.~Srivastava, and R.~Venkatesan, ``Constraint-based invariant inference over predicate abstraction,'' in \emph{Verification, Model Checking, and Abstract Interpretation: 10th International Conference, VMCAI 2009, Savannah, GA, USA, January 18-20, 2009. Proceedings 10}.\hskip 1em plus 0.5em minus 0.4em\relax Springer, 2009, pp. 120--135.

\bibitem{fedyukovich2018accelerating}
G.~Fedyukovich and R.~Bod{\'\i}k, ``Accelerating syntax-guided invariant synthesis,'' in \emph{Tools and Algorithms for the Construction and Analysis of Systems: 24th International Conference, TACAS 2018, Held as Part of the European Joint Conferences on Theory and Practice of Software, ETAPS 2018, Thessaloniki, Greece, April 14-20, 2018, Proceedings, Part I 24}.\hskip 1em plus 0.5em minus 0.4em\relax Springer, 2018, pp. 251--269.

\bibitem{echenim2019ilinva}
M.~Echenim, N.~Peltier, and Y.~Sellami, ``Ilinva: Using abduction to generate loop invariants,'' in \emph{Frontiers of Combining Systems: 12th International Symposium, FroCoS 2019, London, UK, September 4-6, 2019, Proceedings 12}.\hskip 1em plus 0.5em minus 0.4em\relax Springer, 2019, pp. 77--93.

\bibitem{ernst2007daikon}
M.~D. Ernst, J.~H. Perkins, P.~J. Guo, S.~McCamant, C.~Pacheco, M.~S. Tschantz, and C.~Xiao, ``The daikon system for dynamic detection of likely invariants,'' \emph{Science of computer programming}, vol.~69, no. 1-3, pp. 35--45, 2007.

\bibitem{le2019sling}
T.~C. Le, G.~Zheng, and T.~Nguyen, ``Sling: using dynamic analysis to infer program invariants in separation logic,'' in \emph{Proceedings of the 40th ACM SIGPLAN Conference on Programming Language Design and Implementation}, 2019, pp. 788--801.

\bibitem{jhala2006practical}
R.~Jhala and K.~L. McMillan, ``A practical and complete approach to predicate refinement,'' in \emph{International Conference on Tools and Algorithms for the Construction and Analysis of Systems}.\hskip 1em plus 0.5em minus 0.4em\relax Springer, 2006, pp. 459--473.

\bibitem{mcmillan2010lazy}
K.~L. McMillan, ``Lazy annotation for program testing and verification,'' in \emph{Computer Aided Verification: 22nd International Conference, CAV 2010, Edinburgh, UK, July 15-19, 2010. Proceedings 22}.\hskip 1em plus 0.5em minus 0.4em\relax Springer, 2010, pp. 104--118.

\bibitem{gupta2009tests}
A.~Gupta, R.~Majumdar, and A.~Rybalchenko, ``From tests to proofs,'' in \emph{International Conference on Tools and Algorithms for the Construction and Analysis of Systems}.\hskip 1em plus 0.5em minus 0.4em\relax Springer, 2009, pp. 262--276.

\bibitem{barrett2011cvc4}
C.~Barrett, C.~L. Conway, M.~Deters, L.~Hadarean, D.~Jovanovi{\'c}, T.~King, A.~Reynolds, and C.~Tinelli, ``cvc4,'' in \emph{Computer Aided Verification: 23rd International Conference, CAV 2011, Snowbird, UT, USA, July 14-20, 2011. Proceedings 23}.\hskip 1em plus 0.5em minus 0.4em\relax Springer, 2011, pp. 171--177.

\bibitem{barbosa2022cvc5}
H.~Barbosa, C.~Barrett, M.~Brain, G.~Kremer, H.~Lachnitt, M.~Mann, A.~Mohamed, M.~Mohamed, A.~Niemetz, A.~N{\"o}tzli \emph{et~al.}, ``cvc5: A versatile and industrial-strength smt solver,'' in \emph{International Conference on Tools and Algorithms for the Construction and Analysis of Systems}.\hskip 1em plus 0.5em minus 0.4em\relax Springer, 2022, pp. 415--442.

\bibitem{clarke2003counterexample}
E.~Clarke, O.~Grumberg, S.~Jha, Y.~Lu, and H.~Veith, ``Counterexample-guided abstraction refinement for symbolic model checking,'' \emph{Journal of the ACM (JACM)}, vol.~50, no.~5, pp. 752--794, 2003.

\bibitem{zhu2015learning}
H.~Zhu, A.~V. Nori, and S.~Jagannathan, ``Learning refinement types,'' \emph{ACM SIGPLAN Notices}, vol.~50, no.~9, pp. 400--411, 2015.

\bibitem{sharma2013data}
R.~Sharma, S.~Gupta, B.~Hariharan, A.~Aiken, P.~Liang, and A.~V. Nori, ``A data driven approach for algebraic loop invariants,'' in \emph{Programming Languages and Systems: 22nd European Symposium on Programming, ESOP 2013, Held as Part of the European Joint Conferences on Theory and Practice of Software, ETAPS 2013, Rome, Italy, March 16-24, 2013. Proceedings 22}.\hskip 1em plus 0.5em minus 0.4em\relax Springer, 2013, pp. 574--592.

\bibitem{sharma2016invariant}
R.~Sharma and A.~Aiken, ``From invariant checking to invariant inference using randomized search,'' \emph{Formal Methods in System Design}, vol.~48, pp. 235--256, 2016.

\bibitem{garg2016learning}
P.~Garg, D.~Neider, P.~Madhusudan, and D.~Roth, ``Learning invariants using decision trees and implication counterexamples,'' \emph{ACM Sigplan Notices}, vol.~51, no.~1, pp. 499--512, 2016.

\bibitem{ezudheen2018horn}
P.~Ezudheen, D.~Neider, D.~D'Souza, P.~Garg, and P.~Madhusudan, ``Horn-ice learning for synthesizing invariants and contracts,'' \emph{Proceedings of the ACM on Programming Languages}, vol.~2, no. OOPSLA, pp. 1--25, 2018.

\bibitem{yu2023loop}
S.~Yu, T.~Wang, and J.~Wang, ``Loop invariant inference through smt solving enhanced reinforcement learning,'' in \emph{Proceedings of the 32nd ACM SIGSOFT International Symposium on Software Testing and Analysis}, 2023, pp. 175--187.

\bibitem{yao2020learning}
J.~Yao, G.~Ryan, J.~Wong, S.~Jana, and R.~Gu, ``Learning nonlinear loop invariants with gated continuous logic networks,'' in \emph{Proceedings of the 41st ACM SIGPLAN Conference on Programming Language Design and Implementation}, 2020, pp. 106--120.

\bibitem{si2020code2inv}
X.~Si, A.~Naik, H.~Dai, M.~Naik, and L.~Song, ``Code2inv: A deep learning framework for program verification,'' in \emph{Computer Aided Verification: 32nd International Conference, CAV 2020, Los Angeles, CA, USA, July 21--24, 2020, Proceedings, Part II 32}.\hskip 1em plus 0.5em minus 0.4em\relax Springer, 2020, pp. 151--164.

\bibitem{chen2021evaluating}
M.~Chen, J.~Tworek, H.~Jun, Q.~Yuan, H.~P. D.~O. Pinto, J.~Kaplan, H.~Edwards, Y.~Burda, N.~Joseph, G.~Brockman \emph{et~al.}, ``Evaluating large language models trained on code,'' \emph{arXiv preprint arXiv:2107.03374}, 2021.

\bibitem{chowdhery2022palm}
A.~Chowdhery, S.~Narang, J.~Devlin, M.~Bosma, G.~Mishra, A.~Roberts, P.~Barham, H.~W. Chung, C.~Sutton, S.~Gehrmann \emph{et~al.}, ``Palm: Scaling language modeling with pathways. arxiv 2022,'' \emph{arXiv preprint arXiv:2204.02311}, vol.~10, p.~1, 2022.

\bibitem{pei2023can}
K.~Pei, D.~Bieber, K.~Shi, C.~Sutton, and P.~Yin, ``Can large language models reason about program invariants?'' in \emph{International Conference on Machine Learning}.\hskip 1em plus 0.5em minus 0.4em\relax PMLR, 2023, pp. 27\,496--27\,520.

\bibitem{liu2024enhancing}
R.~Liu, G.~Li, M.~Chen, L.-I. Wu, and J.~Ke, ``Enhancing automated loop invariant generation for complex programs with large language models,'' \emph{arXiv preprint arXiv:2412.10483}, 2024.

\bibitem{pirzada2024llm}
M.~A. Pirzada, G.~Reger, A.~Bhayat, and L.~C. Cordeiro, ``Llm-generated invariants for bounded model checking without loop unrolling,'' in \emph{Proceedings of the 39th IEEE/ACM International Conference on Automated Software Engineering}, 2024, pp. 1395--1407.

\end{thebibliography}
